\documentclass{article}[12pt]

\def\usingvc{true}
\IfFileExists{preamble_vc_paper}{\input{preamble_vc_paper}}{}
\usepackage{amsmath,amssymb}
 
\def\eq{\begin{equation}}
\def\en{\end{equation}}
\def\eqa{\begin{eqnarray}}
\def\ena{\end{eqnarray}}
\def\expval#1{\langle \, #1 \,\rangle}
\def\expvalc#1{\expval{#1}_{c}}

\def\partialby#1{\frac{\partial\hfill}{\partial#1}}
\def\opval#1{\langle\langle\, #1 \, \rangle\rangle }
\def\Dop{\mathcal{D}}

\def\Oop{\mathcal{O}}
\def\ord#1{\underline{\mathcal{#1}}}

\begin{document} 
\begin{titlepage}
\def\thepage{}
\ifx\usingvc\undefined{}\else{
}\fi
\begin{center}

\vskip .5in 
{\Large \bf Gradient formula for the beta-function of
2d quantum field theory}


\vskip .5in {\large Daniel Friedan}${}^{1}$ and
{\large Anatoly Konechny}${}^{2}$

\vskip 0.5cm
{${}^{1}$Department of Physics and Astronomy\\
Rutgers, The State University of New Jersey,\\
Piscataway, New Jersey 08854-8019 USA\\
and\\
Natural Science Institute,
The University of Iceland\\
Reykjavik, Iceland\\[0.5ex]
email address: friedan@physics.rutgers.edu\\[1.5ex]
${}^{2}$Department of Mathematics,
Heriot-Watt University,\\
Riccarton, Edinburgh, EH14 4AS, UK\\
and\\
Maxwell Institute for Mathematical Sciences, 
Edinburgh, UK}\\[0.5ex]
email address: anatolyk@ma.hw.ac.uk
\end{center}

\vskip .5in

\begin{center}\today \end{center}
\vskip .2in
 \large
\begin{abstract} 
 We give a non-perturbative proof of a gradient formula 
for beta functions of two-dimensional quantum field theories. The gradient formula has the form 
$ \partial_{i}c = - (g_{ij}+\Delta g_{ij} +b_{ij})\beta^{j} $ where $\beta^{j}$ are the beta functions, 
$c$ and $g_{ij}$ are the Zamolodchikov 
$c$-function and metric, $b_{ij}$ is an antisymmetric tensor introduced by H. Osborn and $\Delta g_{ij}$ 
is a certain metric correction. The formula is derived under the assumption of  stress-energy conservation 
and certain conditions on the infrared behaviour the most significant of which is the condition that the large 
distance limit of the field theory does not exhibit spontaneously broken global conformal symmetry.
Being specialized to non-linear sigma models this formula  
implies  a one-to-one correspondence between renormalization group fixed points and critical points of $c$.
\end{abstract}

\end{titlepage}
\newpage \large
\tableofcontents
\newpage
\section{Introduction} \large
\renewcommand{\theequation}{\arabic{section}.\arabic{equation}}
\setcounter{equation}{0} 
Change of scale in quantum field theories (QFTs) is governed by renormalization group (RG) transformations. 
If a space of theories is parameterized by coupling constants $\{\lambda^{i}\}$ the RG transformations 
are governed by a beta-function vector field 
\begin{equation}\label{beta_fn}
\mu \frac{d\lambda^{i}}{d\mu}  = \beta^{i}(\lambda) \, . 
\end{equation}

The idea that  RG flows  could be gradient flows, that is 
\eq \label{grad_prop}
\beta^{i}(\lambda) = -G^{ij}(\lambda)\frac{\partial S(\lambda)}{\partial \lambda^{j}}
\en
for some metric $G^{ij}(\lambda)$ and  potential function $S(\lambda)$ defined on the theory space, 
has some history. One of the earliest papers devoted to this question was \cite{WZ}. It was suggested in that paper 
that RG flows are gradient flows in a wide variety of situations.  Gradient flows have some special properties. 
Thus, if the metric $G^{ij}$ is positive definite,  the scale derivative 
of the potential function is negative definite
\eq \label{monot}
\mu\frac{dS}{d\mu} = \beta^{i}\frac{\partial S}{\partial \lambda^{i}} = 
-G_{ij}\beta^{i} \beta^{j} \le 0
\en
and therefore $S$ monotonically decreases along the flow. This demonstrates irreversibility of the RG flows and 
forbids limiting cycle behaviour. Another appealing property of gradient flows is that the matrix 
of anomalous dimensions $\partial_{i}\beta^{j}$ is symmetric and thus their eigenvalues 
at critical points, that give critical exponents, are always real.

The first perturbative computations in support 
of this idea were done for four-dimensional theories \cite{WZ2}. Later more evidence was found in the 
context of two dimensional general sigma models \cite{FriedanNM1}, \cite{FriedanNM2}. 
In \cite{FriedanNM2} a gradient formula of the form (\ref{grad_prop}) was formulated for such models and shown to hold 
up to two loops for a particular class of sigma models. A crucial ingredient for a gradient formula for general 
sigma models was the introduction of the dilaton field \cite{FradkinTs1}, \cite{FradkinTs2}. 
 It was shown in \cite{Callanetal1}, \cite{Callanetal2}  
that including the dilaton couplings into a general sigma model one finds that the vanishing beta function equations 
are equivalent to critical points of a certain functional 
at the leading order in  $\alpha'$. A gradient formula of the form  
(\ref{grad_prop}) was checked for general sigma models in \cite{Tseytlin0} to the first two orders in $\alpha'$.
 In string theory conformal  sigma models describe strings propagating on the sigma model 
target manifolds. The sigma model  couplings parameterize  a metric $G_{IJ}$, antisymmetric tensor $B_{IJ}$ 
and a dilaton field $\Phi$  defined on the target space manifold.
The gradient property (\ref{grad_prop}) attains a special significance in this context 
becoming  a manifestation 
of the string action principle. The condition for conformal invariance is that the beta functions vanish: 
$\beta^{G}=\beta^{B}=\beta^{\Phi}=0$.
 It is equivalent 
to string equations of motion. The gradient property (\ref{grad_prop}) thus means that the string equations of motion arise by varying a functional of couplings - 
$S$, which can be identified with the string action functional.

Another reinforcement of the gradient conjecture (\ref{grad_prop}) for two-dimensional theories came 
from the  Zamolodchikov $c$-theorem \cite{Zam}.  The last one is a general theorem applicable to unitary 2D theories 
that states that there is a function $c$ on the space of theories that monotonically decreases along the RG flows and 
coincides with the Virasoro central charge at fixed points. (We give a slightly modified proof of this theorem in section 
\ref{Zam_form}). The theorem was proved by constructing $c$ whose scale derivative takes  the 
form  of the right hand side of (\ref{monot}) with a certain positive definite  metric.  It was natural to conjecture that a 
gradient formula of the form (\ref{grad_prop}) holds with $S$ being the $c$ function and $G_{ij}$ - the   
Zamolodchikov metric.   This was shown to hold at the leading 
order in conformal perturbation theory near fixed points \cite{Zam}, \cite{Zam2}.  
In the context of nonlinear sigma models this idea was discussed in 
\cite{Tseytlin_c}. It was  argued in \cite{Tseytlin_c} that for the purposes  of string theory the $c$-function cannot provide 
a suitable potential function (we comment more on this in section \ref{sec:final}). Other  potential functions for RG flows of nonlinear sigma 
models were considered in \cite{Tseytlin1}, \cite{Tseytlin_c}, \cite{Tseytlin2}, \cite{OsbornNLSM1}, \cite{OsbornNLSM2} which were shown to be related 
to the central charge  and to each other. 
 In   \cite{OsbornNLSM2}  a potential function for nonlinear sigma models was constructed 
  assuming the existence of a sigma model zero mode integration measure with certain properties. 
  It was shown that a measure with the required properties can be constructed infinitesimally but a proof of the integrability of that construction is still lacking. An essential tool proposed in \cite{OsbornNLSM2} for deriving gradient formulas was 
  the use of Wess-Zumino consistency conditions on local Weyl transformations in the presence of curved metric 
  and sources. This technique was applied in \cite{Osborn} to a class of quantum field theories subject to 
  certain power counting restrictions. It was shown that for these theories a  gradient formula holds of 
  a slightly different form than (\ref{grad_prop}) :
  \eq\label{eq:Osborngr0}
  \partial_{i}c =- g_{ij}\beta^{j} - b_{ij}\beta^{j}
  \en
  where $c$ and $g_{ij}$ are  the Zamolodchikov's metric and $c$-function \cite{Zam} and $b_{ij}$ is 
  a certain antisymmetric tensor. The necessity to introduce an antisymmetric tensor along with the 
  Zamolodchikov's metric can be demonstrated by the use of conformal perturbation theory. Thus it was 
  shown in \cite{Friedmanetal} by explicit perturbative calculations 
  that the one-form $g_{ij}\partial^{j} c$ is not closed for some flows\footnote{The obstruction 
   to closedness  occurs  at the next to leading order  in  perturbation. }. Still, as we will explain in the next section, 
  Osborn's gradient formula (\ref{eq:Osborngr0}), although very inspiring, falls short 
  of providing a general gradient formula. The main content of the present work is a derivation of a gradient formula that
    generalizes formula (\ref{eq:Osborngr0})  to a much wider class of theories that includes nonlinear 
    sigma models as well.

To finish the historical overview we mention here that
a general gradient formula was proven for boundary renormalization group flows in two dimensions \cite{FriedanKon}. 
Such flows happen in QFTs defined on a half plane (or a cylinder) when the bulk theory is conformal but the 
boundary condition breaks the conformal invariance.
One of the implications of the boundary gradient formula is  a proof of  Affleck and Ludwig's $g$-theorem \cite{AL} 
which is a statement analogous to Zamolodchikov's $c$-theorem. A string theory interpretation of this  
gradient formula is that it provides an off-shell action for open strings. 
The boundary gradient formula was proved under certain assumptions on the UV behaviour which are 
reminiscent of the power counting restrictions of \cite{Osborn}. Nevertheless we will show in the present paper 
that any assumptions of this kind can be dispensed with in proving a bulk gradient formula. 

The paper is organized as follows. In section 2 after  introducing some notations we explain in more detail Osborn's  
gradient formula (\ref{eq:Osborngr0}) and the assumptions that went into proving it. We then state our  main 
result - a  general gradient formula (\ref{new_gf}) and discuss the assumptions needed to prove it. 
In section 3 we give a proof of Zamolodchikov's formula and recast it in the form that we use as a starting point     
for proving the gradient formula. In section 4 the first steps of the proof are explained. At the end of those steps we 
express the quantity $\partial_{i}c  + g_{ij}\beta^{j} + b_{ij}\beta^{j}$ built from the elements 
present in (\ref{eq:Osborngr0}) via three point functions with a certain  contact operator present in them.
To analyze these three point functions we develop a sources and operations formalism in section 5. A short summary 
of the formalism is provided in subsection 5.2. After discussing the Callan-Symanzik equations in section 6 we resume the 
proof in section 7  putting to use the Wess-Zumino consistency conditions on the local renormalization operation and 
our infrared assumptions.
At the end of section 7  an infrared regulated gradient formula is obtained. In section 8 the proof 
is concluded by removing the infrared cutoff. Section 9 contains a discussion of the properties of the gradient formula 
and the assumptions used in proving it. In section 9.5 the gradient formula is specialized to the nonlinear sigma model 
case and a proof is given of the correspondence between RG fixed points and stationary points of $c$.
In section 10 we conclude with some final remarks. 



\section{The general gradient formula}
 \setcounter{equation}{0}
In this paper we consider two-dimensional  Euclidean quantum field theories equipped 
with a conserved stress-energy tensor $T_{\mu\nu}(x)$. 
The stress-energy tensor measures  the response of the theory to metric perturbations, 
so that if $Z[g_{\mu\nu}]$ is a partition function defined on a 2-dimensional plane with metric 
$g_{\mu\nu}(x)=\delta_{\mu\nu} + \delta g_{\mu\nu}$
\eq \label{Tresp}
\delta \ln Z = \frac{1}{2}\iint\!\! d^{2}x\, \langle \delta g_{\mu \nu}T^{\mu\nu}(x)\rangle \, . 
\en
In two dimensions any metric can be made conformally flat so that  
$g_{\mu\nu}(x) = \mu^{2}(x) \delta_{\mu\nu}$ where the function $\mu(x)$ sets the local scale. 
A change of local scale is generated by the trace of stress-energy tensor $\Theta(x)\equiv g^{\mu\nu}T_{\mu\nu}(x)$ 
\eq\label{scale_tr}
\mu(x)\frac{\delta \ln Z}{\delta \mu(x)} = \langle \Theta(x) \rangle \, . 
\en
For correlation functions computed on ${\mathbb R}^{2}$ with  constant scale $\mu$ the change of 
scale is obtained by integrating over an insertion of $\Theta(x)$
\eq\label{eq:scale}
\mu\frac{\partial}{\partial \mu}  \langle {\cal O}_{1}(x_1)\dots {\cal O}_{n}(x_n)\rangle_{c} =
\int\!\! d^{2}x\,  \langle \Theta(x) {\cal O}_{1}(x_1)\dots {\cal O}_{n}(x_n)\rangle_{c} \, . 
\en
Here ${\cal O}_{1}\, , \dots {\cal O}_{n}$ are local operators and
the subscript $c$ at the correlator brackets marks connected correlators.

Assume that a family of renormalizable QFTs is parameterized by renormalized coupling 
constants $\lambda^{i}$, $i=1,\dots, N$. We assume that an action principle \cite{Schwinger} 
is satisfied. This means that for each  coupling $\lambda^{i}$ there exists a local operator $\phi_{i}(x)$ 
such that for any set of local operators ${\cal O}_{1}\, , \dots ,{\cal O}_{n}$
\begin{eqnarray}\label{ap1}
\frac{\partial}{\partial \lambda^{i}} \langle {\cal O}_{1}(x_1)\dots {\cal O}_{n}(x_n)\rangle_{c} = 
\int\!\! d^{2}x\, \langle \phi_{i}(x) {\cal O}_{1}(x_1)\dots {\cal O}_{n}(x_n)\rangle_{c}\, . 
\end{eqnarray}
Note that the integrability of the integrand in (\ref{eq:scale}),(\ref{ap1}) assumes the appropriate infrared 
behaviour of the correlators. 

Assume further that the couplings $\lambda^{i}$ can be promoted to local sources $\lambda^{i}(x)$ for the fields 
$\phi_{i}(x)$. The generating functional $\ln Z$ then in general depends on the scale factor 
$\mu(x)$ and the sources $\lambda^{i}$, and the action principle (\ref{ap1}) means 
 that in addition to  (\ref{Tresp}) we have 
\eq \label{ap2}
\frac{\delta \ln Z}{\delta \lambda^{i}(x)} = \langle \phi_{i}(x) \rangle \, . 
\en
A correlation function of the form 
\eq \label{gen_corr}
\langle \phi_{i_{1}}(x_1)\phi_{i_{2}}(x_2)\dots \phi_{i_{n}}(x_2)\Theta(y_1)\Theta(y_2)\dots \Theta(y_{m}) \rangle_{c}
\en
 evaluated on a flat ${\mathbb R}^{2}$ can be obtained by taking variational derivatives 
of $\ln Z$ with respect to the sources $\lambda^{i}$ and the metric scale factor $\mu$ and then 
setting the sources and the scale to be constant.
In a renormalized theory the correlators (\ref{gen_corr}) are distributions. They form a basic set of 
local physical quantities defined in a given QFT.  

In a renormalizable QFT a change of scale 
can be compensated by changing the couplings $\lambda^{i}$ according to (\ref{beta_fn}).  
By the action principle (\ref{ap1}) this implies that 
$\Theta(x) = \beta^{i}\phi_{i}(x)$ 
where $\beta^{i}$ are the beta functions. This equation should be understood as an operator equation, 
that is, as an equation that holds inside correlation functions (\ref{gen_corr}) up to contact terms 
(i.e.  up to 
distributions supported on subsets of measure zero). 
The use of sources $\lambda^{i}(x)$ and non-constant Weyl factor $\mu(x)$ facilitates bookkeeping of 
the contact terms.
 In the presence of non-constant 
$\lambda^{i}(x)$ and $\mu(x)$ one can expand  the difference $\Theta(x) - \beta^{i}(\lambda(x))\phi_{i}(x)$ in terms of 
derivatives of the sources and metric \cite{Osborn}. The expansion must by covariant with respect to changes 
of coordinates. This requirement ensures that the contact terms respect the conservation of stress-energy tensor. 
In \cite{Osborn} H. Osborn assumed that this expansion has the form 
\eq\label{OsbornD}
\Theta(x)- \beta^{i}\phi_{i}(x) = \frac{1}{2}\mu^{2}R_{2}(x)C(\lambda) + 
\partial^{\mu}[W_{i}(\lambda)\partial_{\mu}\lambda^{i}] + 
\frac{1}{2}\partial_{\mu}\lambda^{i}\partial^{\mu}\lambda^{j}G_{ij}(\lambda)
\en
where 
\eq
\mu^{2}R_{2}(x) = -2\partial_{\mu}\partial^{\mu}\ln \mu(x)
\en
is the two-dimensional curvature density. 
Note that in (\ref{OsbornD}) $C$, $W_{i}$ and $G_{ij}$ are functions of $\lambda$ evaluated on $\lambda^{i}(x)$ that
 depend on $x$ via $\lambda^{i}(x)$ only.
Effectively equation (\ref{OsbornD}) gives a local version of renormalization group equation.
 Using the Wess-Zumino consistency conditions for the local renormalization 
group  transformations (\ref{OsbornD}) H. Osborn 
derived a gradient formula \cite{Osborn}
\eq\label{Osborn_grad}
\partial_{i}c + g_{ij}\beta^{j}+ b_{ij}\beta^{j}  = 0
\en
where $c$ and $g_{ij}$ are  the Zamolodchikov $c$-function and metric \cite{Zam} 
defined in terms of two-point functions as 
\eq
c = 4\pi^{2}\left (
 x^{\mu}x^{\nu} x^{\alpha}x^{\beta}
- x^{2}g^{\mu\nu}  x^{\alpha}x^{\beta}
- \frac{1}{2}x^{2} x^{\mu}g^{\nu\alpha}x^{\beta}
\right )
{\expvalc{T_{\mu\nu}(x)\,T_{\alpha\beta}(0)}}_{\big /\Lambda|x|=1}
\label{eq:corig}
\en
\eq
g_{ij} = 6\pi^{2}  \Lambda^{-4} \,{\expvalc{\phi_{i}(x)\,\phi_{j}(0)}}_{\big / \Lambda 
|x|=1}
\label{eq:gijorig}
\en
where $\Lambda^{-1}$ is a fixed arbitrary 2-d distance.
The tensor $b_{ij}$ is an antisymmetric two-form that can be expressed as 
\begin{equation}\label{eq:bijorig}
b_{ij}=\partial_{i}w_{j} - \partial_{j}w_{i}\, , \quad w_{i} = 3\pi \int\!\! d^{2}x\,x^{2}\theta(1-\Lambda|x|)
\langle \phi_{i}(x)\Theta(0)\rangle_{c} 
\end{equation}
where $\Lambda$ is the same mass scale used in the definition of $c$ and $g_{ij}$. 
The most restrictive assumption in \cite{Osborn} appears to be the form of expansion (\ref{OsbornD}). 
The fact that the expansion does not go beyond the second order in derivatives suggests a certain power 
counting principle. 
Such a principle could be provided in the vicinity of an ultraviolet  fixed 
point by the standard power counting arguments for renormalizability.
Even with such a counting principle the expansion (\ref{OsbornD}) is too restrictive. Thus it omits   
terms of the form $\partial_{\mu} \lambda^{i}J^{\mu}_{i}(x)$ where $J^{\mu}_{i}(x)$ are local vector fields 
which can be prescribed engineering dimension 1.  Such terms in the scale anomaly can be generated 
by near marginal perturbations near fixed points. In particular they are present in generic current-current 
perturbations of Wess-Zumino-Witten theories \cite{FKinprep}. 
Another class of theories for which (\ref{OsbornD}) is too restrictive is general nonlinear sigma models. 
In this case one needs to allow the quantities $C$, $W_{i}$ and $G_{ij}$ in (\ref{OsbornD}) to have  a 
non-trivial operator content. The case of sigma models was covered separately in \cite{Osborn} (see also 
\cite{Tseytlin1}, \cite{Tseytlin2}, \cite{Tseytlin_c},  \cite{OsbornNLSM1}, \cite{OsbornNLSM2} and references 
therein). It was shown that a gradient formula analogous to (\ref{Osborn_grad}) can be derived provided 
a sigma model integration measure with certain  properties exists. In the present paper we will go 
beyond Osborn's  UV assumptions allowing for an arbitrary local covariant expansion with operator-valued 
coefficients replacing (\ref{OsbornD}). Making instead assumptions about the infrared behaviour we 
derive a general formula
\eq \label{new_gf}
\partial_{i}c + (g_{ij}+\Delta g_{ij}) \beta^{j}+b_{ij}\beta^{j}  = 0 \, .
\en
 The metric correction $\Delta g_{ij}$ is constructed via two point functions of $\phi_{i}$ with 
 the currents $J_{j}^{\mu}(x)$  arising from the expansion generalizing expansion (\ref{OsbornD}) 
 (see formulas (\ref{eq:gLij}), (\ref{eq:deltagij})).
 Alternatively  $\Delta g_{ij}$ can be expressed via  3 point functions with the pure-contact 
field
$D(x) = \Theta(x) -\beta (x)$ (formula (\ref{eq:deltagijreg})). 
 Formula (\ref{new_gf}) is derived under 
two separate  assumptions on the infrared behaviour. The first assumption is that the action principle (\ref{ap1}) holds 
for one and two point functions of operators $\phi_{i}$ that assumes that these functions  are at least once differentiable. This ensures in 
particular that the $c$-function is once differentiable. The second assumption is that for 
any vector field $J_{\mu}(x)$ we have 
\eq\label{IR1}
\lim_{|x|\to \infty} |x|^{3}\langle J_{\mu}(x)T_{\alpha\beta}\rangle_{c}=0 \,. 
\en 
This condition is equivalent to requiring that the the long distance limit of the QFT does not exhibit 
spontaneously broken global conformal symmetry. (Recall that at fixed points special conformal symmetry requires $T(z)$ to decay at infinity as $|z|^{-4}$.) As a simple example in section   \ref{IRexample} demonstrates, this condition is 
essential. If in a scale invariant theory the global conformal symmetry is broken via  boundary conditions at infinity
the value of the central charge may vary with  moduli.

 Our considerations include the nonlinear sigma model case. We thus show that in order 
to have a gradient formula we may replace the somewhat obscure technical assumption on the measure given in 
\cite{OsbornNLSM2} by a more conceptually clear  assumption on the stress-energy tensor behaviour (\ref{IR1}) 
which we show to be a necessary assumption  in section \ref{IRexample}. 
A question remains, of course, how 
one can check whether  our infrared conditions hold in any  given theory. 
Since in the nonlinear sigma model the expectation values of 
diffeomorphism invariant local operators are believed to be free of perturbative infrared divergences
 they must be analytic in the couplings (\cite{Elitzur}, \cite{David}). This means that the first infrared assumption can be controlled in perturbation theory. 
 It is less clear to us whether one can control the infrared behaviour of $T_{\mu\nu}$ perturbatively.
  We are planning to discuss applications of our general result (\ref{new_gf}) to nonlinear sigma models 
  in more details in a separate paper  \cite{FKinprep}.

\section{Zamolodchikov's formula}\label{Zam_form}\label{sect:cgijformulas}
\setcounter{equation}{0}
Zamolodchikov proved in \cite{Zam}  the following formula 
\eq
\mu\frac{\partial c}{\partial{\mu}}  = -\beta^{i}g_{ij}\beta^{j}
\label{eq:zam}
\en
where $\mu$ is the RG scale, $c$ is the $c$-function (\ref{eq:corig}) and 
$g_{ij}$ is the metric introduced in (\ref{eq:gijorig}). This formula 
implies that $c$ decreases under the renormalization group flow
and is stationary exactly at the fixed points.  $c$ is normalized so that  
at   fixed points its value coincides with the value of the Virasoro central charge.

Note that the $c$-function and the metric $g_{ij}$ depend on $\Lambda$ only through the dimensionless ratio 
$\Lambda/\mu$, because according to (\ref{Tresp}) and (\ref{ap1})  the fields $T_{\mu\nu}(x)$ and $\phi_{i}(x)$ 
are densities in $x$,
implying that their 2-point functions take the form
\eqa \label{eq:muscale}
\expvalc{T_{\mu\nu}(x)\,T_{\alpha\beta}(0)} &=& 
\mu^{4} F_{\mu\nu\alpha\beta}(\mu x)\, ,  \nonumber \\
\expvalc{\phi_{i}(x)\,\phi_{j}(0)} &=& \mu^{4} F_{ij}(\mu x)\, , \nonumber \\
\expvalc{T_{\mu\nu}(x)\,\phi_{i}(0)} &=& 
\mu^{4} F_{\mu\nu, i}(\mu x)
\,.
\ena

Before we set out to prove the general gradient formula it is instructive to 
go over a proof of  formula (\ref{eq:zam}). 
One way to prove equation (\ref{eq:zam}) is to derive
alternative formulas for $c$ and $g_{ij}$
\eqa
c&=& - \int\!\! d^{2}x \,\ G_{\Lambda}(x) 
\,\expvalc{\Theta(x)\,\Theta(0)}
\label{eq:c}\\
g_{ij}&=& -\Lambda\partialby\Lambda  \int\!\! d^{2}x \, G_{\Lambda}(x) \,
\expvalc{\phi_{i}(x)\,\phi_{j}(0)}
\label{eq:gij}
\ena
where
\eq
G_{\Lambda}(x) = 3\pi  x^{2} \theta(1-\Lambda |x|)
\,.
\en
These are the formulas for $c$ and $g_{ij}$ 
that we will use in the proof of the gradient formula.
Equation (\ref{eq:zam}) follows immediately
from formulas (\ref{eq:c}) and (\ref{eq:gij}):
\eqa
\mu\frac{\partial c}{\partial \mu}  = -\Lambda\frac{\partial c}{ \partial \Lambda} &=&
\Lambda\partialby\Lambda \int\!\! d^{2}x \,\ G_{\Lambda}(x) 
\,\expvalc{\Theta(x)\,\Theta(0)} \nonumber \\
 &=&  \int\!\! d^{2}x \,\ \Lambda\frac{\partial G_{\Lambda}(x) }{\partial \Lambda} 
\,\expvalc{\beta^{i}\phi_{i}(x)\,\beta^{j}\phi_{j}(0)} \nonumber \\ 
 &=& - \beta^{i} g_{ij} \beta^{j}
\,.
\ena
Replacing $\expvalc{\Theta(x)\,\Theta(0)}$ by 
$\expvalc{\beta^{i}\phi_{i}(x)\,\beta^{j}\phi_{j}(0)}$ in the second line is allowed because they 
differ only by a contact term in $x$, which gives no contribution
since the smearing function
$\Lambda\partial G_{\Lambda}(x)/\partial 
\Lambda $ is supported away from $x=0$.


While formula (\ref{eq:gij}) is evidently equivalent to 
 formula (\ref{eq:gijorig}) the equivalence of formulas (\ref{eq:corig}) and (\ref{eq:c}) for $c$ 
 is shown as follows.
 Combine
the special identity in two space-time dimensions
\eq
\left (
x^{2}g^{\mu\nu}g^{\alpha\beta}
-g^{\mu\nu}x^{\alpha}x^{\beta}
-x^{\mu}x^{\nu}g^{\alpha\beta}
+ 2 g^{\mu\alpha}x^{\nu}x^{\beta}
- x^{2}g^{\mu\alpha}g^{\nu\beta}
\right )
\expvalc{T_{\mu\nu}(x)T_{\alpha\beta}(0)} = 0
\en
with the Ward identity
\eq
\partial^{\mu}\expvalc{T_{\mu\nu}(x)T_{\alpha\beta}(0)} = 0
\en
and CPT invariance
\eq
\expvalc{T_{\mu\nu}(x)T_{\alpha\beta}(0)} = 
\expvalc{T_{\mu\nu}(-x)T_{\alpha\beta}(0)}
=\expvalc{T_{\alpha\beta}(x)T_{\mu\nu}(0)}
\en
to calculate
\eq \label{int_rel}
\partial^{\mu}
\left [
\left ( 2 x^{\nu}x^{\alpha}x^{\beta}
-2 x^{2} x^{\nu } g^{\alpha\beta} 
- x^{2} g^{\nu\alpha} x^{\beta}
\right )
\expvalc{T_{\mu\nu}(x)T_{\alpha\beta}(0)}
\right ]
= -3 x^{2} \expvalc{\Theta (x) \Theta(0)} \, .
\en
It follows from (\ref{int_rel}) that
\begin{align}
- \int d^{2}x \,&\ G_{\Lambda}(x) 
\,\expvalc{\Theta(x)\,\Theta(0)}\nonumber \\
&= \pi
\int d^{2}x \,\theta(1-\Lambda|x|)
\,\partial^{\mu}
\left [
\left ( 2 x^{\nu}x^{\alpha}x^{\beta}
-2 x^{2} x^{\nu } g^{\alpha\beta} 
- x^{2} g^{\nu\alpha} x^{\beta}
\right )
\expvalc{T_{\mu\nu}(x)T_{\alpha\beta}(0)}
\right ] \nonumber \\
&= \pi
\int d^{2}x \,\delta(1-\Lambda|x|)
|x|^{-2} x^{\mu}
\left ( 2 x^{\nu}x^{\alpha}x^{\beta}
-2 x^{2} x^{\nu } g^{\alpha\beta} 
- x^{2} g^{\nu\alpha} x^{\beta}
\right )
\expvalc{T_{\mu\nu}(x)T_{\alpha\beta}(0)}
\nonumber \\
&=2\pi^{2} \left (  2 x^{\mu}x^{\nu}x^{\alpha}x^{\beta}
-  x^{2} x^{\mu}x^{\nu } g^{\alpha\beta} 
- x^{2}  g^{\mu\nu } x^{\alpha}x^{\beta }
- x^{2} x^{\mu}g^{\nu\alpha} x^{\beta}
\right )
{\expvalc{T_{\mu\nu}(x)T_{\alpha\beta}(0)}}
_{\big / \Lambda |x|=1}
\end{align}
which demonstrates the equivalence of (\ref{eq:corig}) and 
(\ref{eq:c}).

\section{The proof of the gradient formula (first steps)}
\setcounter{equation}{0}

We start by defining a 1-form $r_{i}$ by the equation
\eq
\partial_{i}c + g_{ij}\beta^{j} +b_{ij}\beta^{j} +r_{i} = 0
\en
and show that the remainder term $r_{i}$
can be expressed in terms of correlation functions of $\Theta(x)$ and 
$\phi_{i}(x)$ with the pure-contact field
$D(x) = \Theta(x) -\beta (x)$.Infrared behavior of the correlation functions will be an important 
issue,
so we introduce an IR cutoff at $|x|=L\gg \Lambda^{-1}$ and keep track 
of the error terms.
Our assumptions about IR behavior will be designed to ensure
the vanishing of the IR error in the limit 
$L\rightarrow \infty$.

We start out by recasting $g_{ij}\beta^{j}$ as
\begin{equation}
g_{ij}\beta^{j} = 6\pi^{2} \Lambda^{-4}\langle \phi_{i}(x)\phi_{j}(0)\beta^{j}\rangle_{\big / \Lambda |x|=1} 
= 6\pi^{2}\Lambda^{-4}\langle \phi_{i}(x)\Theta(0)\rangle_{\big / \Lambda |x|=1} 
\end{equation}
which is valid because $\beta^{j}\phi_{j}(0)$ differs from $\Theta(0)$ only by contact terms.
This can be further rewritten as 
\begin{eqnarray}
g_{ij}\beta^{j} = -\Lambda\frac{\partial}{\partial \Lambda} \int\!\!d^{2}x\, 
G_{\Lambda}(x)\langle \phi_{i}(x) \Theta(0)\rangle_{c} 
= \mu\frac{\partial}{\partial \mu} \int\!\!d^{2}x\, 
G_{\Lambda}(x)\langle \phi_{i}(x) \Theta(0)\rangle_{c} 
\end{eqnarray}
where the scaling property (\ref{eq:muscale}) was   used on the last step.
Finally using (\ref{eq:scale}) we obtain 
\eq \label{formula1}
g_{ij}\beta^{j} =
\int d^{2}y\int d^{2}x \; G_{\Lambda}(x) \,
\expvalc{\Theta(y)\,\phi_{i}(x)\,\Theta(0)} \, .
\en
Formula (\ref{formula1}) is infrared safe but as we want to impose the IR cutoff systematically,  we write instead
\eq
g_{ij}\beta^{j}
+ E_{1}
= \int_{|y|<L} d^{2}y \int d^{2}x \; G_{\Lambda}(x) \,
\expvalc{\Theta(y)\,\phi_{i}(x)\,\Theta(0)}
\label{eq:gijbetaj}
\en
The Ward identity gives the error term
\eqa
E_{1} &=&  \int_{|y|<L} d^{2}y \;\partial^{\mu}
\left [ y^{\nu} \int d^{2}x \; G_{\Lambda}(x) \,
\expvalc{T_{\mu\nu}(y)\,\phi_{i}(x)\,\Theta(0)}\right ]
\nonumber \\
&=& 2\pi\,y^{\mu}y^{\nu}
\int d^{2}x \; G_{\Lambda}(x) \,{\expvalc{T_{\mu\nu}(y)\,\phi_{i}(x)\,\Theta(0)}}_{\big /|y|=L}
\label{eq:E1}
\ena
which certainly vanishes in the limit $L\rightarrow\infty$.

We next turn our attention to the derivative $\partial_{i}c$. 
Assuming that  $c$ can be differentiated with respect to the coupling constants 
$\lambda^{i}$, we can write using formula (\ref{eq:c}) for $c$ and the action principle (\ref{ap1})
\eq
\partial_{i} c = - \int d^{2}y \, 
\int d^{2}x \,\ G_{\Lambda}(x) 
\,\expvalc{\phi_{i}(y) \,\Theta(x)\,\Theta(0)}
\,.
\en
Again, we regularize in the IR as
\eq
\partial^{L}_{i} c  = - \int_{|y|<L} d^{2}y \, 
\int d^{2}x \,\ G_{\Lambda}(x) 
\,\expvalc{\phi_{i}(y) \,\Theta(x)\,\Theta(0)}
\label{eq:ci} \, . 
\en
Formulas (\ref{eq:gijbetaj}) and (\ref{eq:ci}) can be combined 
to get
\eqa
&&\partial^{L}_{i} c + g_{ij}\beta^{j} +E_{1} =
\int_{|y|<L} d^{2}y
\int d^{2}x \, G_{\Lambda}(x) \,
\expvalc{\Theta(y)\,\phi_{i}(x)\,\Theta(0)-\phi_{i}(y) 
\,\Theta(x) \,\Theta(0)} \nonumber \\
&&=
\int_{|y|<L} d^{2}y
\int d^{2}x \, G_{\Lambda}(x) \,
\expvalc{
\left [ \beta(y)+D(y)\right ]\,\phi_{i}(x)\,\Theta(0)
-\phi_{i}(y) \,\left [\beta(x)+D(x)\right ]
\,\Theta(0)} \nonumber \\
&&= -b^{L}_{ij}\beta^{j}
+ \int_{|y|<L} d^{2}y
\int d^{2}x \, G_{\Lambda}(x) \,
\expvalc{
D(y)\,\phi_{i}(x)\,\Theta(0)
-\phi_{i}(y) \,D(x) \,\Theta(0)}
\label{eq:riderived}
\ena
where we have introduced the 2-form $b^{L}_{ij}$
\eq
b^{L}_{ij} =
\int_{|y|<L} d^{2}y \int d^{2}x \, G_{\Lambda}(x) 
\,\expvalc{\phi_{i}(y) \,\phi_{j}(x) \,\Theta(0)
-\phi_{j}(y) \,\phi_{i}(x) \,\Theta(0)}
\,.
\label{eq:bLij}
\en
Equation (\ref{eq:riderived}) can be written as
\eq
\partial^{L}_{i}c + g_{ij}\beta^{j} +E_{1}+ b^{L}_{ij}\beta^{j} +r^{L}_{i} = 0 
\label{eq:gradformula1}
\en
with
\eq
r^{L}_{i} = \int_{|y|<L} d^{2}y
\int d^{2}x \, G_{\Lambda}(x) \,
\expvalc{\phi_{i}(y) \,D(x) \,\Theta(0)
-
D(y)\,\phi_{i}(x)\,\Theta(0)}
\label{eq:ri}
\en

Equations (\ref{eq:gradformula1}), (\ref{eq:ri}) are the main results of this section. We will later show 
that under our assumptions on the infrared behaviour the limits 
\begin{equation}
\partial_{i}c=\lim_{L\to \infty}\partial_{i}^{L}c \, , \qquad 
b_{ij} = \lim_{L\to \infty}b_{ij}^{L}
\end{equation} 
exist. The error term $E_{1}$ goes to zero as $L\to \infty$. 
The remainder term $r_{i}^{L}$ is expressed via correlation functions 
involving the pure-contact field $D(x)$. 
In order to investigate this term 
we develop a sources and operations formalism for calculating correlation functions
of  $D(x)$.


\section{Sources and operations}
\setcounter{equation}{0}
In this section we present a general formalism that allows computing correlation functions of 
pure contact fields using functional differential operators acting on functionals of sources and 
metric. The general exposition is somewhat tedious so for the reader's convenience  we present the most 
important ingredients necessary to understand the proof of the gradient formula in a separate subsection \ref{subsec:summary}. 

\subsection{General formalism}
So far we have introduced the fields $\phi_{i}(x)$ as operators conjugate to the coupling 
constants $\lambda^{i}$ that parameterize a renormalizable 2D QFT. It will be convenient 
to assume  that the set $\phi_{i}$ is  complete in a given class of fields which we denote by ${\cal F}$. 
The class of fields  
 can be a complete 
set of spin-0 relevant and near marginal fields. We could define such fields without a reference to 
a particular  fixed point by requiring that the corresponding coupling constant belongs to 
some family of renormalizable theories with finitely many couplings (there are finitely many couplings 
for which $\beta^{i}$ is not identically zero). This will not work for the nonlinear sigma models, for which 
the set of  couplings is infinite, but 
in that case we could talk about near-relevant and near-marginal couplings using the  engineering scaling 
dimensions introduced via  free fields. As yet another possibility we could  assume that the set $\{\phi_{i}\}$  
spans all spin-0 local fields and work with a Wilsonian RG. We will keep the class of fields ${\cal F}$ unspecified 
throughout this section assuming only that ${\cal F} $ is closed under RG the precise sense of which we
 will discuss below. 
 In general a field $O(x)$ is defined via its distributional correlation functions with other fields. 
If $O(x)\in {\cal F}$ the completeness of $\{\phi_{i}\}$ means that there are unique coefficients $O^{i}$ 
such that the field $O(x) - O^{i}\phi_{i}(x)$ has vanishing correlation functions with all fields from ${\cal F}$ 
inserted away from $x$. The field $O(x) - O^{i}\phi_{i}(x)$ is thus a pure contact field, that is its correlation 
functions are distributions supported on a subset of measure zero in $x$. We can define {\it ordinary} fields $O(x)$ as 
fields for which the correlations of  $O(x) - O^{i}\phi_{i}(x)$ are zero as distributions. This means that the 
distributional correlation functions of such fields are obtained from those of the fields $\phi_{i}(x)$  by 
contracting them with the appropriate coefficients $O^{i}$. 

Whatever ${\cal F}$ we choose it is essential that the 
trace of stress-energy tensor can be expanded in these fields: $\Theta(x) = \beta^{i}(\lambda)\phi_{i}(x)$. 
It is worth noting that the set $\phi_{i}$ may include total derivative fields. Although the correlation 
functions are independent of the corresponding coupling constants the beta functions may be non-trivial and 
total derivatives may thus contribute to $\Theta(x)$. 
Let us further introduce sources $\lambda^{i}(x)$ for all  fields $\phi_{i}(x)$ so that the generating functional 
$\ln Z$ depends on these sources and the metric scale factor $\mu(x)$ with equations (\ref{scale_tr}) and 
(\ref{ap2}) satisfied.  
This means that $\phi_{i}(x)$ and $\Theta(x)$ are represented by functional derivatives 
\eq\label{func_der}
\phi_{i}(x) = \frac{\delta }{\delta \lambda^{i}(x)} \, , \qquad \Theta(x) = \mu(x) \frac{\delta}{\delta \mu(x)}
\en
which we chose to denote by the same symbols. 
The action of these functional derivatives on $\ln Z$ generates distributional correlation functions (\ref{gen_corr}). 
To facilitate the use of differential operators in computing correlation functions we introduce a shorthand notation 
\eqa 
\rangle\rangle &=& \ln Z \\
\langle\langle &=& \mbox{restriction of functionals to constant sources and flat 2-d metric}
\ena
so
\eq
\opval{\phi_{i_{1}}(x_{1})\cdots \Theta(y_{1})\cdots } = 
\expvalc{\phi_{i_{1}}(x_{1})\cdots \Theta(y_{1})\cdots}
\en
where, on the left hand side, the $\phi_{i}(x)$ and $\Theta(x)$ are functional 
differential operators (\ref{func_der}), while on the right hand side they are  fields.

Define {\it operations}  ${\cal O}(x)$ to be first order local differential operators acting on functionals of the 
sources and 2-d metric. The word local here means that the coefficients of the functional derivatives in 
an operation  given at $x$ can depend only on the values of $\lambda(x)$, $\mu(x)$ and finitely many 
derivatives thereof. 
An ordinary field $O(x)=O^{i}(\lambda)\phi_{i}(x)$ is naturally 
assigned an operation ${\cal O}(x)=O^{i}(\lambda(x))\phi_{i}$. Operations of this form we will call ordinary. 
An arbitrary operation $\Oop (x)$ gives rise  to an ordinary field denoted $\ord{O}(x)$ via
\eq\label{eq:ordinaryf}
\expvalc{\ord{O}(x) \,\phi_{i_{1}}(x_{1})\cdots \Theta(y_{1})\cdots } 
= \opval{\Oop(x)\, \phi_{i_{1}}(x_{1})\cdots \Theta(y_{1})\cdots }\, . 
\en
Although the above formula specifies distributional 
correlation functions containing only a single $\ord{O}(x)$ 
it defines uniquely the coefficients ${O}^{i}$ in   $\ord{O}(x)={O}^{i}\phi_{i}(x)$ 
and thus in principle fixes the correlation functions containing arbitrarily many
$\ord{O}(x)$. The ordinary operation ${O}^{i}\phi_{i}(x)$ corresponding to $\ord{O}(x)$ will be denoted 
by the same symbol $\ord{O}(x)$. Define  pure-contact operations ${\cal O}(x)$ as operations satisfying   
$\ord{O}(x) = 0$, i.e.,
\eq
\langle\langle\, \Oop (x) = 0\, 
\en
Then
\eqa\label{eq:purecontactcorr}
\opval{\phi_{i_{1}}(x_{1})\cdots \Theta(y_{1})\cdots {\cal O}(0)}
&=& \opval{\lbrack \phi_{i_{1}}(x_{1}),\,\Oop (0)\rbrack \cdots \Theta(y_{1})\cdots }
+ \cdots \nonumber 
\\&&\qquad{}
+ \opval{ \phi_{i_{1}}(x_{1})\cdots \lbrack 
\Theta(y_{1}),\,\Oop (0)\rbrack \cdots}
+ \cdots
\ena
is a sum of contact terms.

We would like now to construct an operation for a given 
operator that can be used in computing its correlation functions from $\ln Z$. 
Since we know how to do this for ordinary operators it suffices to solve this problem 
for a pure contact field. Let $O(x)\in {\cal F}$ be a pure contact field that does 
not explicitly depend on $\lambda^{i}$ that is $[\partial_{i},O(x)]=0$. 
Then we can construct a pure contact operation $ {\tilde O}(x)$ by requiring 
\eq\label{pc1}
\opval{\phi_{i_{1}}(x_{1})\cdots \Theta(y_{1})\cdots {\tilde O}(x)} = 
\langle \phi_{i_{1}}(x_{1})\cdots \Theta(y_{1})\cdots O(x)\rangle_{c} \, .
\en
This essentially fixes ${\tilde O}(x)$  because in physical correlators singularities appear only 
when some of the insertions coincide. The only ambiguity in ${\tilde O}(x)$ is operations 
annihilating $\ln Z$. Any choice however suffices for practical purposes. 
With this definition given an arbitrary operator $A(x)\in {\cal F}$ its correlators 
with the fundamental fields $\phi_{i_{k}}(x_{k})$, $\Theta(y_{l})$ can be computed using 
the ordinary operation $\ord{A}(x)=A^{i}\phi_{i}$ and the contact operation 
\eq\label{calA}
{\cal A}(x)\equiv \widetilde{[A-\ord{A}]}(x)
\en according to 
\eq\label{pc2}
\langle \phi_{i_{1}}(x_{1})\cdots \Theta(y_{1})\cdots A(x)\rangle_{c} = 
\opval{\phi_{i_{1}}(x_{1})\cdots \Theta(y_{1})\cdots [\ord{A}(x) +  {\cal A}(x)]} \, . 
\en
In the above correlation function the contact terms proportional to $\delta(x-x_{i_{k}})$ are 
essentially fixed by the action principle (\ref{ap1}). The extra contributions arising from 
the explicit dependence of the coefficients $A^{i}$ on $\lambda^{j}$'s are accounted for 
  by commuting the operation $ \ord{A}(x)$ to the left. Similarly 
the contact terms proportional to $\delta(x-y_{i_{k}})$ are fixed by the change of scale equation (\ref{eq:scale}). 
All contact term contributions proportional to derivatives of delta functions  are obtained 
by  commuting the pure contact  operation ${\cal A}(x)$ to the left until it annihilates $\langle\langle$. 



Consider now the operator $\Theta(x)$. Assuming, as we agreed before, that $\Theta(x) = \beta^{i}\phi_{i}\equiv \beta(x)$ 
in the operator sense means that $\underline{\Theta}=\beta(x)$ and the field 
$D(x)=\Theta(x)-\beta(x)$ is pure contact.  As $\Theta(x)$ does not explicitly depend on $\lambda^{i}$
($\mu\partial/\partial\mu$ and $\partial/\partial \lambda^i$ commute) we can define a pure contact operation 
 $\Dop(x)$ in accordance with the general rule (\ref{calA}), (\ref{pc2}). The field $\Theta(x)$ is special in that 
 it is represented by a variational derivative (\ref{func_der}). This implies that 
 \eq\label{ren_oper}
 \opval{\phi_{i_{1}}(x_{1})\cdots \Theta(y_{1})\cdots [\Theta(x) - \beta(x) - \Dop(x)]}=0
 \en  
 that can be written more succinctly as a first order functional differential equation on the generating functional 
 \eq\label{eq:RGopgen}
 [\Theta(x) - \beta(x) - \Dop(x)] \ln Z = 0 \, . 
 \en
Knowing the pure contact operation $\Dop(x)$ the correlation functions of $D(x)$ with
any number of $\phi_{i}(x)$ and $\Theta(x)$ can be calculated as
\eqa\label{Dcorrs}
&& \expvalc{D(x) \,\phi_{i_{1}}(x_{1}) \dots \Theta(y_{1})\dots} =  
\opval{(\Theta(x)-\beta(x)) \,\phi_{i_{1}}(x_{1}) \dots \Theta(y_{1})\cdots } \nonumber \\
&&=
\opval{[(\Theta(x)-\beta(x)),\,\phi_{i_{1}}(x_{1}) \dots  \Theta(y_{1})\dots]  }
+\opval{ \phi_{i_{1}}(x_{1}) \dots  \Theta(y_{1})\dots\Dop(x)} \nonumber \\
&&=
\opval{ [\phi_{i_{1}}(x_{1}) \dots \Theta(y_{1})\dots,\, (\Dop(x)+\beta(x)-\Theta(x))]} \nonumber \\
&&=
\opval{ [\phi_{i_{1}}(x_{1}) ,\, \Dop(x)] \dots \Theta(y_{1})\dots} + \dots 
+ \opval{ \phi_{i_{1}}(x_{1})\dots [\Theta(y_{1}),\Dop(x)]\dots } + \dots \nonumber \\
&&+ \partial_{i_{1}}\beta^{i} \delta (x-x_{1}) \expvalc{ \phi_{i}(x_{1}) \dots \Theta(y_{1})\cdots} + \dots
\ena
where equation (\ref{ren_oper}) was used on the second line, $\langle\langle \Dop(x)=0$ was used on the 
third line and 
\eq
\lbrack\phi_{i_{1}}(x_{1}),\, \beta(x) ] = \delta(x-x_{1})\partial_{i_{1}}\beta^{i} \phi_{i}(x_{1})
\,.
\en
was used on the last line.

The form of $\Dop(x)$ is constrained by 2-d covariance and locality. In general it can be written as an expansion 
in derivatives of the sources $\lambda^{i}$ and covariant derivatives of the curvature with coefficients 
being ordinary operations. It is interesting to consider additional restrictions on $\Dop(x)$ from 
power counting rules. We will distinguish two such rules which we call a {\it loose  power counting} and a 
{\it strict power 
counting}. In both cases the expansion of $\Dop(x)$ goes only up to two derivatives in the sources and metric. 
In the loose power counting rule the coefficients can have a nontrivial operator content. Explicitly 
in this case we can write  
\eq
\Dop(x) =
\frac12 \mu^{2} R_{2}(x) C(x) + \partial_{\mu}\lambda^{i}(x) J^{\mu}_{i}(x)
+\partial^{\mu}\left [ W_{i}(x)\partial_{\mu}\lambda^{i}\right ]
+\frac12 \partial_{\mu}\lambda^{i}\partial^{\mu}\lambda^{j} G_{ij}(x) 
\label{eq:Doploose}
\en
where $C(x)$, $W_{i}(x)$, $G_{ij}(x)$ are ordinary spin-0 fields, and 
$J^{\mu}_{i}(x)$ is an ordinary spin-1 field,
and where the 2-d curvature is given by
$$
 \mu^{2}R_{2}(x) = -2\partial^{\mu}\partial_{\mu} \ln \mu(x)\, . 
$$
Two comments are in order here. Firstly, notice the appearance of vector fields 
$J^{\mu}_{i}(x)$ in the expansion. As we defined operations only for spin-zero fields 
to accommodate fields and operations of nontrivial spin we need to introduce new 
fundamental fields and new sources for those fields. 
 While used to obtain distributional correlation functions involving 
operators of nontrivial spin such sources are always set to zero in the end of a computation.
The operation $\Dop(x)$  does contain terms proportional to the tensorial sources and their derivatives.
However our proof avoids using 
the explicit form of such  terms and 
we will not introduce the tensor field sources  explicitly not to clutter the computations.
Nevertheless the operations like  $J^{\mu}_{i}(x)$, when appear,  should be understood in this sense. 

Secondly, notice that in the power counting scheme used the operators $C(x)$, $W_{i}(x)$, $G_{ij}(x)$ 
must have dimension near zero. This means that, using the fixed point language, we allow for slightly 
irrelevant terms to appear in $\Dop(x)$. This is a common consideration used for general nonlinear sigma models 
\cite{FriedanNM2}. The loose power counting thus accommodates perturbative nonlinear sigma models. 

If one assumes the UV behaviour is governed by a unitary fixed point,  the only dimension zero 
operator is the identity,  and the total UV dimension of 
$\Dop(x)$ must be strictly 2 then the operators $C(x)$, $W_{i}(x)$, $G_{ij}(x)$ must be all proportional to 
the identity operator. We call this restrictions a strict power counting rule. 
It applies in a vicinity of a unitary fixed point that has a  discrete spectrum of conformal dimensions.
Under the additional assumption 
that there are no operators $J^{\mu}_{i}(x)$ appearing in $\Dop(x)$ the case of the strict power
 counting was investigated in \cite{Osborn}.

 Finally the case when the only restrictions on $\Dop(x)$ come from the general covariance and locality 
 can be referred to as Wilsonian. We will prove the general gradient formula (\ref{new_gf}) in the Wilsonian case. 
 The  proof is simplified if we impose loose power counting. We will be discussing in parallel how our steps look 
 in that case.

As a last comment in this section note that due to equation (\ref{eq:RGopgen}) the operation $\Dop(x)$ is subject to Wess-Zumino consistency conditions 
\eq
[\Theta(x) - \beta(x) - \Dop(x),\,  \Theta(y) - \beta(y) - \Dop(y)] \ln Z = 0 
\en
which will be exploited in sections \ref{sect:DTheta}, \ref{sect:baregrad}.





\subsection{Summary} \label{subsec:summary}
Operations ${\cal O}(x)$ 
are local first order differential operators defined on functionals of the sources $\lambda^{i}(x)$ 
and metric. For the fundamental fields $\phi_{i}(x)$ and the trace of stress energy tensor $\Theta(x)$ the 
corresponding operations are the functional derivatives (\ref{func_der}).
We introduced the notation $\langle\langle {\cal O}_{1}(x_1)\dots {\cal O}(x_n) \rangle\rangle$ 
for a sequence of operations ${\cal O}_{i}(x_{i})$ applied to the generating functional $\ln Z \equiv \rangle\rangle$ 
with the result  restricted to constant sources and metric 
(the restriction is signified by the symbol $\langle\langle\, $).

Given an operation ${\cal O}(x)$ one can extract a field from it by restricting it to constant sources and metric (\ref{eq:ordinaryf}).  
The resulting fields are denoted $\underline{{\cal O}}(x)$ and are called ordinary fields. Such fields have the form 
$\underline{{\cal O}}(x) = O^{i}\phi_{i}(x)$. 
A pure contact operation is an operation ${\cal O}(x)$ for which $\underline{{\cal O}}(x)=0$.

For ordinary fields the distributional correlation 
functions are completely fixed by those of the fields $\phi_{i}$. More generally a given field $A(x)$ equals a linear combination 
of fundamental fields: $A(x) = A^{i}\phi_{i}(x)$ only up to contact terms. Such contact terms can be stored in a pure contact operation ${\cal A}(x)$ according to (\ref{pc2}). For the trace of stress-energy tensor $\Theta(x)$ 
we have $\Theta(x)=\beta^{i}\phi_{i}(x)\equiv \beta(x)$ up to contact terms. The corresponding contact terms 
 are stored in a pure contact operation $\Dop(x)$. The generating functional satisfies an equation 
 $[\Theta(x) - \beta(x) - \Dop(x)]\ln Z = 0$ which can be used to compute correlation functions involving 
 the field $D(x) = \Theta(x) - \beta(x)$ according to (\ref{Dcorrs}). The form of $\Dop(x)$ is 
 constrained by locality and general covariance. It can be further constrained by a power counting principle. 
 We distinguish a strict power counting, which applies to  a vicinity of a unitary fixed point 
 with discrete spectrum of conformal dimensions,  and a loose power counting that is suitable for 
 describing renormalizable nonlinear sigma models. For the loose power counting case $\Dop(x)$ can be explicitly 
 written as in formula (\ref{eq:Doploose}).


\section{The Callan-Symanzik equations}
\setcounter{equation}{0}
In the operations formalism the Callan-Symanzik equations for correlators involving 
 fields $\phi_{i}(x)$ and $\Theta(y)$ can be obtained by integrating equation (\ref{Dcorrs}) 
 over $x$:
 \begin{eqnarray}\label{CSeq1}
&& \left ( \mu\partialby\mu - \beta^{i}\partialby{\lambda^{i}} \right )
\expvalc{\phi_{i_{1}}(x_{1})\dots \Theta(y_{1})\dots  } =\int\!\! d^{2}x\, \opval{
D(x)\,
\phi_{i_{1}}(x_{1})\dots \Theta(y_{1})\dots
}
\nonumber \\
&&=
\partial_{i_{1}}\beta^{i}\expvalc{\phi_{i}(x_{1})\dots \Theta(y_{1})\dots } 
+ \int\!\!d^{2}x\, \opval{
[\phi_{i_{1}}(x_{1}), \Dop(x)]\dots \Theta(y_{1})} +  
\dots \nonumber \\
&& +  \int\!\!d^{2}x\,\opval{ \phi_{i_{1}}(x_{1})\dots [\Theta(y_{1}),\Dop(x)]\dots } + \dots 
\end{eqnarray}

It is convenient to define the following  operations
\eqa\label{eq:calDops}
\Dop\phi_{i}(x) &=& \int\!\!d^{2}y\,  \lbrack\phi_{i}(x),\, \Dop(y)]\, , \nonumber \\
\Dop\Theta(x) &=& \int\!\! d^{2}y\, \lbrack\Theta(x),\, \Dop(y)]
\,.
\ena
In view of (\ref{CSeq1}) the operations $\Dop\phi_{i}(x)$ and $\Dop\Theta(x)$ can be 
interpreted
as extra contributions 
to the Callan-Symanzik equations.

We further notice that 
\eq\label{constrs}
\int d^{2}x\,\langle\langle\,\Dop \Theta(x) =0 \, , \qquad 
\int d^{2}x\,\langle\langle\,\Dop \phi_{i}(x) = 0\, . 
\en
This follows from the fact that 
$\int d^{2}x\,\Theta(x) = \mu\partial/\partial\mu$, $\int 
d^{2}y\,\phi_{i}(x) =\partial/\partial\lambda^{i}$, and every term in $\Dop(y)$  
is proportional to derivatives of $\lambda^{i}(y)$ and $\mu(y)$.
Equations (\ref{constrs}) imply that
 there must be ordinary spin-1 fields (and ordinary operations respectively) 
 $J^{\mu}(x)$ and $J_{i}^{\mu}(x)$ 
such that
\eq
\underline{\Dop \Theta}(x) = - \partial_{\mu} J^{\mu}(x)\, , 
\en
\eq \label{Jimu}
\underline{\Dop \phi_{i}}(x) = - \partial_{\mu} J_{i}^{\mu}(x)
\,.
\en

If we impose loose power counting, so that $\Dop(x)$ is given by 
equation (\ref{eq:Doploose}), then
\eqa
\Dop\Theta(x) &=& -\partial^{\mu}\partial_{\mu} C(x)\\
\Dop\phi_{i}(x) &=& -\partial_{\mu}\left [
J_{i}^{\mu}(x) + \partial^{\mu}\lambda^{j} G_{ij}(x)
\right ] + \partial_{\mu} \lambda^{j}\partial_{i} J_{j}^{\mu}(x) + 
\frac{1}{2}\partial_{\mu}\lambda^{j}\partial^{\mu}\lambda^{k}\partial_{i}G_{jk}(x)
\ena
so
\eqa
J^{\mu}(x) &=& \partial^{\mu} C(x) 
\ena
and $J_{i}^{\mu}(x)$ defined in  (\ref{Jimu}) in general (without any power counting assumptions) 
coincides with the 
 coefficient in 
the expansion of $\Dop(x)$ based on loose power counting, 
equation (\ref{eq:Doploose}).
In general (without any power counting restrictions)  since all terms in 
$\Dop(x)$ are proportional to derivatives of 
the sources and/or to derivatives of $\mu(x)$ there exists a scalar operator $C(x)$ such that 
$J^{\mu}(x)=\partial^{\mu}C(x)$.

The Callan-Symanzik equations (\ref{CSeq1}) for the correlation functions at 
non-coincident points  (neglecting contact terms) can be now be written as 
\begin{align} \label{eq:CS1}
\mu\partialby\mu
\expvalc{\phi_{i_{1}}(x_{1})\dots\Theta(y_{1})\dots }
&=
\beta^{i}\partialby{\lambda^{i}}
\expvalc{\phi_{i_{1}}(x_{1})\cdots\Theta(y_{1})\dots }
+\expvalc{\Gamma\phi_{i_{1}}(x_{1}) \dots \Theta(y_{1})\dots} 
+ \dots    \nonumber \\
&\qquad\qquad{}+\expvalc{\phi_{i_{1}}(x_{1}) \dots 
[-\partial_{\mu}J^{\mu}(y_{1})]\dots} 
+ \dots
\end{align}
where
\eq \label{eq:CS2}
\Gamma\phi_{i_{1}}(x_{1})
= \partial_{i_{1}}\beta^{i}\phi_{i}(x_{1})-\partial_{\mu}J_{i_{1}}^{\mu}(x_{1})
\,.
\en
The terms involving the beta functions can be put into the Lie derivative ${\cal L}_{\beta}$ so that 
equation (\ref{eq:CS1}) takes a more succinct form 
\begin{eqnarray} \label{eq:CS3}
&&[\mu\partialby\mu - {\cal L}_{\beta}]
\expvalc{\phi_{i_{1}}(x_{1})\cdots\Theta(y_{1})\dots }
 \nonumber \\
&& =\expvalc{[-\partial_{\mu}J_{i_{1}}^{\mu}(x_{1})] \dots \Theta(y_{1})\dots} 
+ \dots    
+\expvalc{\phi_{i_{1}}(x_{1}) \dots 
[-\partial_{\mu}J^{\mu}(y_{1})]\dots} 
+ \dots
\end{eqnarray}


\section{The proof continued}
\setcounter{equation}{0}
We now come back to the proof of the gradient formula which we left at the end 
of section 4. We express the remainder term $r^{L}_{i}$ of equation (\ref{eq:ri})
in the source-operation formalism.
The 3-point functions occurring in equation~\ref{eq:ri} can be written as
\eqa
\expvalc{\phi_{i}(y)\,D(x)\,\Theta(0)} &=&
\opval{\phi_{i}(y)\, \Theta(0) \, [\Theta(x)-\beta(x)]}
+\partial_{i}\beta^{j} \delta^{2}(y-x)\opval{\Theta(0)\,\phi_{j}(x)}\nonumber \\
&=&
\opval{\phi_{i}(y)\, \Theta(0)\, \Dop(x)}
+\partial_{i}\beta^{j} \delta^{2}(y-x)\opval{\Theta(0)\,\phi_{j}(x)}\, , 
\\
\expvalc{D(y)\,\phi_{i}(x)\,\Theta(0)} &=&
\opval{\phi_{i}(x)\, \Theta(0) \, \Dop(y)}
+\partial_{i}\beta^{j} \delta^{2}(x-y)\opval{\Theta(0)\,\phi_{j}(y)}\, , 
\ena
so
\eq
\expvalc{\phi_{i}(y) \,D(x) \,\Theta(0)
-
D(y)\,\phi_{i}(x)\,\Theta(0)}
=
\opval{\phi_{i}(y) \,\Theta(0)\,\Dop(x)
-
\phi_{i}(x)\,\Theta(0)\,\Dop(y)}\, . 
\en
Substituting the last relation in equation (\ref{eq:ri}) and
using $\langle\langle\,  \Dop(x) =0$,
gives
\eqa\label{riLint}
r^{L}_{i} &=& \int_{|y|<L} d^{2}y
\int\!\! d^{2}x \, G_{\Lambda}(x) \,
\opval{\phi_{i}(y) \,\Theta(0)\,\Dop(x)
-
\phi_{i}(x)\,\Theta(0)\,\Dop(y)} \nonumber\\
&=& \int_{|y|<L} d^{2}y
\int\!\! d^{2}x \, G_{\Lambda}(x) \,
\opval{\phi_{i}(y) \,\lbrack \Theta(0), \, \Dop(x)\rbrack
+\lbrack\phi_{i}(y), \,\Dop(x)\rbrack  \,\Theta(0)} \nonumber \\
&&
-\int_{|y|<L} d^{2}y
\int\!\! d^{2}x \, G_{\Lambda}(x) \,
\opval{\phi_{i}(x) \,\lbrack \Theta(0), \, \Dop(y)\rbrack
+\lbrack\phi_{i}(x), \,\Dop(y)\rbrack \, \Theta(0)} \, . 
\ena
Note that $\Dop(x)$ is a pure-contact 
operation,
and $|x|\le \Lambda^{-1}\ll L$,
so that
\eqa
\int_{|y|<L} d^{2}y\; \lbrack \Theta(0), \, \Dop(y)\rbrack &=& 
\int\!\! d^{2}y\; \lbrack \Theta(0), \, \Dop(y)\rbrack
= \Dop\Theta(0)\\
\int_{|y|<L} d^{2}y\; \lbrack\phi_{i}(x)\,\Dop(y)\rbrack &=&
\int\!\! d^{2}y\; \lbrack\phi_{i}(x)\,\Dop(y)\rbrack = 
\Dop\phi_{i}(0)\\
\int_{|y|<L} d^{2}y\; \langle\langle\, \lbrack\phi_{i}(y), \,\Dop(x)\rbrack &=&
\int\!\! d^{2}y\; \langle\langle\, \lbrack\phi_{i}(y), \,\Dop(x)\rbrack = 
\lbrack\partial_{i}, \,\Dop(x)\rbrack = 0
\,.
\ena
Using these relations in (\ref{riLint})  we obtain
\begin{align}
r^{L}_{i}
= -  \int_{|y|<L} d^{2}y\,
&12\pi\opval{\phi_{i}(y) \,C_{2}(0)}
- \int d^{2}x \, G_{\Lambda}(x) \,
\opval{\phi_{i}(x) \,\Dop\Theta(0)}\nonumber \\
&{}- \int d^{2}x \, G_{\Lambda}(x) \,
\opval{\Dop\phi_{i}(x)\,\Theta(0)}
\label{eq:ri2}
\end{align}
where we have defined an operation
\eq
C_{2}(y) = - \int\!\! d^{2}x \,\frac14 x^{2} \,\lbrack \Theta(y), \, \Dop(x)\rbrack
\,.
\en
If loose power counting is imposed,
 $\Dop(x)$ is given by equation (\ref{eq:Doploose}),
and we have
\eq
C_{2}(y) =  - \int\!\! d^{2}x \,\frac14 x^{2} \,
\left [ -\partial^{\mu}\partial_{\mu}\delta^{2}(x-y)
\right ] C(x) = C(y)
\,.
\en
Thus, with loose power counting,
\eq
\Dop\Theta(x) = -\partial^{\mu}\partial_{\mu}C_{2}(x)
\,.
\label{eq:C2C}
\en

We separate $r^{L}_{i}$ into two parts
\eqa
r^{L}_{i} &=&  r^{L}_{i,1}+r^{L}_{i,2}\label{eq:rLiseparated}\\
r^{L}_{i,1} &=& - \int d^{2}x \, G_{\Lambda}(x) \,
\opval{\Dop\phi_{i}(x)\,\Theta(0)} \label{eq:rLi1}\\
r^{L}_{i,2} &=& -  \int_{|y|<L} d^{2}y\,
12\pi \opval{\phi_{i}(y) \,C_{2}(0)}
- \int d^{2}x \, G_{\Lambda}(x) \,
\opval{\phi_{i}(x) \,\Dop\Theta(0)}
\label{eq:rLi2}
\ena
then investigate each in turn. 


\subsection{The IR condition and the sum-rule}\label{sect:IRsum}
We  investigate $r_{i,1}$ first. Our goal is to show that under certain assumptions this 
quantity is proportional to the beta functions.

We have
\eq
\langle\langle\, \Dop\phi_{i}(x) = \langle\langle\, \underline{\Dop\phi_{i}}(x)
= \langle\langle\, \left [ -\partial_{\mu}J_{i}^{\mu}(x) \right ]
\en
so
\eq
\opval{\Dop\phi_{i}(x)\,\Theta(0)}
=
- \opval{\partial_{\mu}J_{i}^{\mu}(x)\,\Theta(0)} =
- \expvalc{\partial_{\mu}J_{i}^{\mu}(x)\,\Theta(0)}\, . 
\en
Substituting this expression in equation (\ref{eq:rLi1}) we get
\eq
r^{L}_{i,1}
=
\int d^{2}x \, G_{\Lambda}(x) \,
\expvalc{\partial_{\mu}J_{i}^{\mu}(x)\,\Theta(0)}
\,.
\label{eq:rLi12}
\en
Now we use the technique similar to the one we used in  the proof of Zamolodchikov's formula 
(see section \ref{sect:cgijformulas}). 
It is straightforward to check that the Ward identity for $T_{\mu\nu}(x)$ implies
\eq
x^{2} \expvalc{\partial_{\mu}J_{i}^{\mu}(x)\,\Theta(0)}
=
\partial_{\mu}
\left [
x^{2} \expvalc{J_{i}^{\mu}(x)\,\Theta(0)} 
-2 
x_{\alpha}x^{\beta}\expvalc{J_{i}^{\alpha}(x)\,T_{\beta}^{\mu}(0)} 
+x^{2}\expvalc{J_{i}^{\alpha}(x)\,T_{\alpha}^{\mu}(0)} 
\right ]
\en
which allows us to perform the integral in equation (\ref{eq:rLi12}), 
obtaining
\eq
r^{L}_{i,1}
=
6\pi^2 \, x_{\mu}
\left [
x^{2} \expvalc{J_{i}^{\mu}(x)\,\Theta(0)} 
-2 
x_{\alpha}x^{\beta}\expvalc{J_{i}^{\alpha}(x)\,T_{\beta}^{\mu}(0)} 
{}+x^{2}\expvalc{J_{i}^{\alpha}(x)\,T_{\alpha}^{\mu}(0)} 
\right ] _{\big / \Lambda|x| = 1}\, . 
\en
 What we want however is an expression proportional to $\beta^{i}$.
Recall that
\eq
G_{\Lambda}(x) =  3\pi    x^{2}\,\theta(1-\Lambda|x|) 
\en
so that 
\eq
G_{0}(x) =  3\pi x^{2}
\label{eq:G0}
\en
and
\eq
G_{0}(x) - G_{\Lambda}(x) = 3\pi  x^{2}\,\theta(\Lambda|x|-1) 
\,.
\en
We write
\eqa
r^{L}_{i,1}
&=&
E_{2}
+\int_{|x|\le L} d^{2}x \, [G_{\Lambda}(x)-G_{0}(x)] \,
\expvalc{\partial_{\mu}J_{i}^{\mu}(x)\,\Theta(0)}\nonumber \\
&=&
E_{2}
+
\int_{|x|\le L} d^{2}x \, [G_{\Lambda}(x)-G_{0}(x)] \,
\expvalc{\partial_{\mu}J_{i}^{\mu}(x)\,\phi_{j}(0)}\beta^{j}
\label{eq:rLi1final}
\ena
with
\eqa
E_{2} &=& \int_{|x|\le L} d^{2}x \, G_{0}(x) \,
\expvalc{\partial_{\mu}J_{i}^{\mu}(x)\,\Theta(0)}\\
&=&6\pi^2 \,  x_{\mu}
\left [
x^{2} \expvalc{J_{i}^{\mu}(x)\,\Theta(0)} 
-2 
x_{\alpha}x^{\beta}\expvalc{J_{i}^{\alpha}(x)\,T_{\beta}^{\mu}(0)} 
{}+x^{2}\expvalc{J_{i}^{\alpha}(x)\,T_{\alpha}^{\mu}(0)} 
\right ] _{\big / |x| = L} \, .
\label{eq:E2}
\ena
We are allowed to replace $\Theta(0)$ with $\beta^{j}\phi_{j}(0)$ to 
obtain equation (\ref{eq:rLi1final})
because $G_{\Lambda}(x)-G_{0}(x)$ vanishes for $\Lambda |x|\le 1$,
so contact terms in the 2-point function make no difference.

The IR error term $E_{2}$ will vanish in the limit $L\rightarrow\infty$ if
the 2-point functions 
$\expvalc{J_{i}^{\mu}(x)\,T_{\alpha\beta}(0)}$
go to zero at large $x$ faster than $|x|^{-3}$:
\begin{equation}\label{eq:IRcond}
\lim\limits_{|x|\to \infty} |x|^{3}\expvalc{J_{i}^{\mu}(x)\,T_{\alpha\beta}(0)}=0 \, .
\end{equation}
A violation of this IR decay condition would mean that
the long distance limit of the quantum field theory 
exhibits spontaneously broken 
global conformal symmetry. Our main IR assumption is that such a 
spontaneous breaking does not take place and equation (\ref{eq:IRcond}) is satisfied.

The condition $\lim_{L\rightarrow \infty} E_{2} =0$ 
is equivalent to the sum rule
\eq \label{eq:IRsum}
\int d^{2}x \, x^{2} \,
\expvalc{\partial_{\mu}J_{i}^{\mu}(x)\,\Theta(0)}
= 0
\,.
\en
Such a sum rule holds for any spin-1 field,
given our infrared  assumption.  


\subsection{The  term $r^{L}_{i,2}$}

Similarly to (\ref{eq:rLi1final}) we want to write $r^{L}_{i,2}$ as an integral over $\Lambda 
|x|>1$ of an expression proportional to $\beta^{j}$.
Equation (\ref{eq:C2C}), which one obtains when the loose power counting is 
imposed,
motivates the following manipulation of 
equation (\ref{eq:rLi2}).
Write the first term, using
equation (\ref{eq:G0}) for $G_{0}(y)$,
\eq
- \int_{|y|<L} d^{2}y\, 12\pi
\opval{\phi_{i}(y) \,C_{2}(0)}
=  - \int_{|y|<L} d^{2}y\,\left [\partial_{\mu}\partial^{\mu}G_{0}(y)\right ] 
\opval{\phi_{i}(y) \,C_{2}(0)}
\en
then integrate by parts.  
Equation (\ref{eq:rLi2}) becomes
\eq
r^{L}_{i,2} = E_{3} - \int_{|y|<L} d^{2}y\,G_{0}(y)
\opval{\phi_{i}( y) \,\partial_{\mu}\partial^{\mu} C_{2}(0)}
- \int d^{2}x \, G_{\Lambda}(x) \,
\opval{\phi_{i}(x) \,\Dop\Theta(0)}
\label{eq:rLi211}
\en
where $E_3$ is an infrared error 
\eq
E_{3}
=
- \int_{|x|<L} d^{2}x \, 
\partial_{\mu}\left [ \partial^{\mu}G_{0}(x)\,\opval{\phi_{i}(0) \,C_{2}(x)}
-G_{0}(x)\,\partial^{\mu} \opval{\phi_{i}(0) \,C_{2}(x)} \right ]
\,.
\label{eq:E3}
\en
We further rewrite equation (\ref{eq:rLi211}) as
\eqa
r^{L}_{i,2} = E_{3} +E_{4}
+ \int_{|x|<L} d^{2}x \, \left [ G_{0}(x)-G_{\Lambda}(x) \right ] \,
\opval{\phi_{i}(x) \,\Dop\Theta(0)}
\label{eq:rLi22}
\ena
where
\eq
E_{4} =
- \int_{|x|<L} d^{2}x \, 
G_{0}(x) \opval{\phi_{i}(x) \,
[\partial_{\mu}\partial^{\mu}C_{2}(0)+\Dop\Theta(0)]} \, . 
\label{eq:E4}
\en
The term $E_{4}$  is identically zero if we assume loose power counting,
by equation (\ref{eq:C2C}).
We will show in section \ref{sect:E4} that in general $E_4$ vanishes as $L\to \infty$. 

In equation (\ref{eq:rLi22}), the integration variable $x$ is bounded 
away from $0$, so we can substitute
\eq
\opval{\phi_{i}(x) \,\Dop\Theta(0)}
= \opval{\Dop\Theta(0)\, \phi_{i}(x)}
= \expvalc{-\partial_{\mu}J^{\mu}(0) \, \phi_{i}(x)}
\en
giving
\eq
r^{L}_{i,2}
= E_{3} +E_{4}
+ \int_{|x|<L} d^{2}x \, \left [ G_{\Lambda}(x) - G_{0}(x)\right ] \,
\expvalc{\phi_{i}(x)  \, \partial_{\mu}J^{\mu}(0)} \, . 
\en
Finally, we will now show that
\eq
\partial_{\mu}J^{\mu}(0) = \beta^{j} \partial_{\mu}J_{j}^{\mu}(0)
\en
so that $r^{L}_{i,2}$ also becomes proportional to $\beta^{j}$, up to IR 
errors,
\eq
r^{L}_{i,2}
= E_{3} +E_{4}
+ \int_{|x|<L} d^{2}x \, \left [ G_{\Lambda}(x) - G_{0}(x)\right ] \,
\expvalc{\phi_{i}(x)  \, \partial_{\mu}J_{j}^{\mu}(0)}\beta^{j}
\,.
\label{eq:rLi2final}
\en




\subsection{The identity $\partial_{\mu}J^{\mu}(x) = \beta^{j}\partial_{\mu}J_{j}^{\mu}(x) $}
\label{sect:DTheta}

We want to show that the ordinary field
\eq
K(x) = \beta^{j}\partial_{\mu}J_{j}^{\mu}(x) -\partial_{\mu}J^{\mu}(x)
\en
is zero, which is to say that all its non-coincident correlation 
functions vanish:
\eq \label{eq:Kcontact}
\expvalc{K(x)\, \phi_{i_{1}}(x_{1}) \dots}=0
\qquad x\ne x_{1},\ldots
\en
In the source/operation formalism, this means that
\eq
\opval{K(x) \, \phi_{i_{1}}(x_{1}) \dots}=0
\qquad x\ne x_{1},\ldots
\en
To show this we first argue that (\ref{eq:Kcontact}) is equivalent to showing that 
\eq\label{eq:K1}
\lbrack \Dop,\, D(x) \rbrack \,\rangle\rangle
= \mathcal{K}_{1}(x) \,\rangle\rangle
\en
for some pure-contact operation $\mathcal{K}_{1}(x)$. We then demonstrate that (\ref{eq:K1}) is 
a consequence of  the Wess-Zumino consistency conditions on $\Dop(x)$. 

It follows from (\ref{eq:calDops}) that
\eqa\label{eq:step1}
\langle\langle\, K(x) 
&=& \langle\langle\,\left [ 
-\partial_{\mu}J^{\mu}(x)+ \beta^{j}\partial_{\mu}J_{j}^{\mu}(x)\right ] \nonumber\\
&=& \langle\langle\,\left [ 
\Dop\Theta(x)-\beta^{j}\Dop\phi_{j}(x) \right ] \nonumber \\
&=& \langle\langle\,\lbrack \Dop,\, \Theta(x)-\beta^{j}\phi_{j}(x) \rbrack \nonumber\\
&=& \langle\langle\,\lbrack \Dop,\, D(x) \rbrack
\ena
where $D(x)= \Theta(x)-\beta^{j}\phi_{j}(x)$ is acting here as 
an operation.
This last calculation implicitly uses the obvious identity
\eq
\langle\langle\,\beta^{j}(\lambda(x)) = \beta^{j}(\lambda) \langle\langle
\en
and its direct implication
\eq\label{eq:betaDcomm}
\langle\langle\,\lbrack\Dop,\, \beta^{j}(\lambda(x))\rbrack =
- \langle\langle\,\beta^{j}(\lambda(x))\, \Dop
=
-\beta^{j}(\lambda) \langle\langle\, \Dop
= 0
\,.
\en
Now we have
\eq
\opval{K(x) \, \phi_{i_{1}}(x_{1}) \cdots} =
\opval{ \lbrack \Dop,\, D(x) \rbrack
\, \phi_{i_{1}}(x_{1}) \cdots}
\,.
\en
The operation $\lbrack \Dop,\, D(x) \rbrack$ commutes with all the 
$\phi_{i_{r}}(x_{r})$ because $x\ne x_{r}$, so
\eq
\opval{K(x) 
\, \phi_{i_{1}}(x_{1}) \cdots} =
\opval{\phi_{i_{1}}(x_{1}) \cdots \lbrack \Dop,\, D(x) \rbrack} 
\,.
\en
We now need to show that
\eq
\opval{\phi_{i_{1}}(x_{1}) \cdots \lbrack \Dop,\, D(x) \rbrack} 
=0 
\qquad x\ne x_{1},\ldots
\en
which by (\ref{eq:purecontactcorr}) equivalent to (\ref{eq:K1}).

Equation (\ref{eq:K1}) follows from the Wess-Zumino consistency conditions.
Recall that we have an equation
\eq\label{eq:WZ1}
0 = [D(x)-\Dop(x)] \,\rangle\rangle \, . 
\en
The Wess-Zumino consistency conditions are
\eq
\lbrack D(x)-\Dop(x), \, D(y)-\Dop(y) \rbrack \,\rangle\rangle=0 \, . 
\en
It follows from 
\eq\label{eq:3comms}
[\Theta(x),\Theta(y)]=0\, , \quad [\Theta(x),\beta(y)]=0\, , \quad [\beta(x),\beta(y)]=0 
\en
that 
\eq
[D(x),D(y)]=0 
\en and therefore (\ref{eq:WZ1}) is equivalent to 
\eq\label{eq:KK1}
\lbrack\Dop(y),\, D(x) \rbrack\,\rangle\rangle
=  - \left ( 
\lbrack D(y), \, \Dop(x)  \rbrack
+\lbrack \Dop(x), \, \Dop(y) \rbrack
\right )\,\rangle\rangle \, . 
\en
The operation $\lbrack \Dop(x), \, \Dop(y) \rbrack$ is evidently pure-contact. It also follows 
from (\ref{eq:scale}) and (\ref{eq:betaDcomm}) that 
\eq
\lbrack \int d^{2}y\,D(y), \, \Dop(x)  \rbrack
\en
is a pure contact operation. Thus
 integrating equation (\ref{eq:KK1}) with respect to $y$ gives
\eq
\lbrack \Dop,\, D(x) \rbrack \,\rangle\rangle
= \mathcal{K}_{1}(x) \,\rangle\rangle
\en
where
\eq
\mathcal{K}_{1}(x) =
- 
\lbrack \int d^{2}y\,D(y), \, \Dop(x)  \rbrack
-\lbrack \Dop(x), \, \Dop \rbrack
\en
is pure contact.
This completes the proof that at
all non-coincident correlation functions of
$\beta^{j}\partial_{\mu}J_{j}^{\mu}(x) -\partial_{\mu}J^{\mu}(x)$
are identically zero.
Therefore\footnote{It is worth noting that relation (\ref{eq:operrel}) is a generalization  
of the Curci-Paffuti relation \cite{CP} known for nonlinear sigma models. By methods similar to those employed 
in this section one can actually prove a stronger relation: $J^{\mu}(x) = \beta^{j}J_{j}^{\mu}(x)$. We do 
not need this stronger relation in the proof of the gradient formula. }
\eq\label{eq:operrel}
\partial_{\mu}J^{\mu}(x) = \beta^{j}\partial_{\mu}J_{j}^{\mu}(x)
\,.
\en

\subsection{ $E_{4}$ is an IR error term}\label{sect:E4} 
We owe a proof that the term $E_4$ given by 
\begin{equation} \label{mainE4}
E_4 = -\int\limits_{|x|<L}\!\! d^2x\, G_{0}(x)\langle\langle \phi_i(x) [
\partial_{\mu}\partial^{\mu} C_2(0) + {\cal D}\Theta(0)]\rangle\rangle
\end{equation}
is an infrared error, that is it vanishes as $L\to \infty$.
The argument is a bit tedious, so the  reader might want to skip this subsection at the first reading.

We have 
Note  that in general (without the assumption of loose power counting)  
we have 
\begin{equation}
[\Theta(0),{\cal D}(y)]= -\partial_{\mu}\partial^{\mu}\delta(y)\,  C_2(y) + 
\partial_{\mu}\partial_{\nu}\partial_{\gamma}\delta(y)\, C_{3}^{\mu \nu \gamma}(y) + \dots 
\end{equation}
where the omitted terms contain derivatives of delta functions of order 4 and higher. 
For our purposes this expansion can be written more compactly as 
\begin{equation}\label{exp}
[\Theta(0),{\cal D}(y)]= -\partial_{\mu}\partial^{\mu}\delta(y)\,  C_2(y) + 
\partial_{\mu}\partial_{\nu}\partial_{\gamma}\delta(y)\, \tilde C_{3}^{\mu \nu \gamma}(y)  
\end{equation}
where $\tilde C_{3}^{\mu \nu \gamma}(y)$ is some  tensor operation. 
Formula (\ref{exp}) implies 
\begin{equation}
\partial_{\mu}\partial^{\mu} C_2(0) + {\cal D}\Theta(0) = \partial_{\mu}\partial_{\nu}\partial_{\gamma} 
\tilde C_{3}^{\mu \nu \gamma}(0) 
\end{equation}
and therefore 
\eq\label{eq:E4ax}
\langle\langle \phi_i(x)[
\partial_{\mu}\partial^{\mu} C_2(0) + {\cal D}\Theta(0)]\rangle\rangle = 
\langle  \partial_{\mu}\partial_{\nu}\partial_{\gamma} 
 \underline{\tilde C_{3}}^{\mu \nu \gamma}(0)\phi_i(x)\rangle  + 
 \langle\langle [\phi_{i}(x),  \partial_{\mu}\partial_{\nu}\partial_{\gamma} 
\tilde C_{3}^{\mu \nu \gamma}(0) ]   \rangle\rangle \, . 
\en
The second term on the right hand side of (\ref{eq:E4ax}) vanishes because it is proportional to a one 
point function of a total derivative operator. 
Thus we obtain
\begin{equation}\label{eq:E4final}
E_4 = -3\pi \int\limits_{|x|<L}\!\! d^2x\, x^2 \langle \phi_i(x) \partial_{\mu}\partial_{\nu}\partial_{\gamma} 
 \underline{\tilde C_{3}}^{\mu \nu \gamma}(0)\rangle 
\end{equation}
which exhibits that $E_4$ is a linear combinations of two point functions at separation $L$. 
Assuming that $\langle \phi_i(L) \underline{\tilde C_{3}}^{\mu \nu \gamma}(0)\rangle $ is integrable at 
infinity (which is consistent with $\langle \underline{\tilde C_{3}}^{\mu \nu \gamma}(0)\rangle =0$ 
being independent of $\lambda_i$) all combinations of two point functions entering $E_4$ go to zero 
as $L\to \infty$. 


\section{Conclusion of the proof}
\setcounter{equation}{0}
Combining our results for $r^{L}_{i,1}$ and  $r^{L}_{i,2}$,
equations (\ref{eq:rLi1final}) and (\ref{eq:rLi2final}),
and substituting in equation (\ref{eq:rLiseparated}),
we get
\eq
r^{L}_{i} 
= E_{2}+E_{3}+E_{4}
+ \left (\Delta g^{L}_{ij}\right ) \beta^{j}
\label{eq:ri4}
\en
with
\eq
\Delta g^{L}_{ij}
=
\int_{|x|<L} d^{2}x \, [G_{\Lambda}(x)-G_{0}(x)] \,
\expvalc{
\phi_{i}(x)\,\partial_{\mu}J_{j}^{\mu}(0)
+
\,\phi_{j}(x)\, \partial_{\mu}J_{i}^{\mu}(0)
}\, . 
\label{eq:gLij}
\en
The metric correction $\Delta g^{L}_{ij}$ can be also written without any direct reference 
to currents $J_{i}^{\mu}(x)$ using the Callan-Symanzik equations (\ref{eq:CS3}) 
\begin{eqnarray}\label{eq:deltagijreg}
\Delta g^{L}_{ij}=  \int_{|x|<L} d^{2}x \, [G_{\Lambda}(x)-G_{0}(x)]
({\cal L}_{\beta} - \mu\frac{\partial}{\partial \mu})\langle \phi_{i}(x)\phi_{j}\rangle  \,
\end{eqnarray}
which, using (\ref{eq:scale}), (\ref{ap1}),   can be written in terms of integrated three point functions of
 fundamental operators up to IR error terms. 

Equation (\ref{eq:gradformula1}) becomes, finally,
the IR-regulated gradient formula
\eq \label{eq:IRreggrad}
\partial^{L}_{i}c + (g_{ij}+\Delta g^{L}_{ij} +
b^{L}_{ij})\beta^{j} + E(L) = 0 
\en
with total error
\eq
E(L) = E_{1}+E_{2}+E_{3}+E_{4}
\,.
\en
The $L$-dependent constituents of the formula are:
 \eq
\partial^{L}_{i} c  = - \int_{|y|<L} d^{2}y \, 
\int d^{2}x \,\ G_{\Lambda}(x) 
\,\expvalc{\phi_{i}(y) \,\Theta(x)\,\Theta(0)}\, , 
\en
\eq
b^{L}_{ij} =
\int_{|y|<L} d^{2}y \int d^{2}x \, G_{\Lambda}(x) 
\,\expvalc{\phi_{i}(y) \,\phi_{j}(x) \,\Theta(0)
-\phi_{j}(y) \,\phi_{i}(x) \,\Theta(0)}\, , 
\en
\eq
E_{1} = 2\pi y^{\mu}y^{\nu}
\int d^{2}x \; G_{\Lambda}(x) \,{\expvalc{T_{\mu\nu}(y)\,\phi_{i}(x)\,\Theta(0)}}_{\big /|y|=L}\, , 
\en
\eq
E_{2} = 6\pi^{2} \,  x_{\mu}
\left [
x^{2} \expvalc{J_{i}^{\mu}(x)\,\Theta(0)} 
-2 
x_{\alpha}x^{\beta}\expvalc{J_{i}^{\alpha}(x)\,T_{\beta}^{\mu}(0)} 
{}+x^{2}\expvalc{J_{i}^{\alpha}(x)\,T_{\alpha}^{\mu}(0)} 
\right ] _{\big / |x| = L}\, , 
\en
\eq
E_{3}
=
- \int_{|x|<L} d^{2}x \, 
\partial_{\mu}\left [ \partial^{\mu}G_{0}(x)\,\opval{\phi_{i}(0) \,C_{2}(x)}
-G_{0}(x)\,\partial^{\mu} \opval{\phi_{i}(0) \,C_{2}(x)} \right ]\, , 
\en
\eq
E_{4} =-3\pi \int\limits_{|x|<L}\!\! d^2x\, x^2 \langle \phi_i(x) \partial_{\mu}\partial_{\nu}\partial_{\gamma} 
 \underline{\tilde C_{3}}^{\mu \nu \gamma}(0)\rangle 
\en
and $\Delta g_{ij}^{L}$ is given in (\ref{eq:gLij}) (see equations (\ref{eq:ci}), (\ref{eq:bLij}), 
(\ref{eq:E1}), (\ref{eq:E2}), (\ref{eq:E3}), (\ref{eq:E4final})).

Now that the infrared regulated formula (\ref{eq:IRreggrad}) is derived we can study its $L\to \infty$ limit. 
Let us recapitulate our assumptions on the infrared behavior. Firstly,  we assume that the action principle holds at least for one and two point functions 
so that the one and two-point functions are at least once differentiable. Secondly, the infrared behavior of the stress-energy tensor correlators 
should satisfy (\ref{eq:IRcond}). The first assumption means that 2,3 and 4-point functions involving 
$\phi_{i}(x)$ or $T_{\mu\nu}(x)$ decay faster than $x^{2}$ when $|x| \to \infty$. 
This together with formula (\ref{eq:IRcond}) imply that
\begin{eqnarray} \label{eq:IRlimits}
&& \lim\limits_{L\to \infty} E(L)=0 \, , \nonumber \\
&& \lim\limits_{L\to \infty} \partial^{L}_{i}c = \partial_{i} c \, , \nonumber \\
&&   \lim\limits_{L\to \infty} b_{ij}^{L}  = b_{ij} 
\end{eqnarray}
where $b_{ij}$ is given by Osborn's 
formula\footnote{The 2-form $b_{ij}$ is exact provided $w_{j}$ defined in (\ref{eq:bijorig}) is 
 differentiable. If one relaxes the differentiability assumptions 
there is room for the limit $b_{ij}= 
\lim_{L\rightarrow\infty}b^{L}_{ij}$ to exist without
$w_{j}$ being differentiable, in which case $b_{ij}$ would be closed 
but not exact. The failure  of differentiability of $w_{j}$ could come from some non-perturbative effects.} (\ref{eq:bijorig}). Note that in showing (\ref{eq:IRlimits}) formula 
(\ref{eq:IRcond}) is needed only to argue that $E_2$ vanishes at infinity while the first infrared assumption 
alone suffices to show all other limits.

 Note that although the same set of assumptions implies 
\begin{equation}
 \lim\limits_{L\to \infty} \Delta g_{ij}^{L} \beta^{j} < \infty 
 \end{equation}
there is no guarantee that the $L\to \infty$ limit of $\Delta g_{ij}^{L}$ is finite. However, 
Infrared divergences, if present in  $\Delta g_{ij}^{L}$, 
are orthogonal to the beta function. Therefore they can be  subtracted to obtain a finite 
quantity $\Delta g_{ij}$ so that the following gradient formula holds
\begin{equation}\label{eq:finalgradf}
\partial_{i}c = -(g_{ij} + \Delta g_{ij} + b_{ij}) \beta^{j} \, 
\end{equation} 
where
\begin{equation}\label{eq:deltagij}
\Delta g_{ij} =  \lim\limits_{L\to \infty} [\Delta g_{ij}^{L} - \mbox{ subtractions }]\, .
\end{equation}
This completes the derivation of the general gradient formula. 



\section{Discussion}
\setcounter{equation}{0}

\subsection{Contact term ambiguities  and scale dependence} \label{contterms}
As the proof of the gradient formula uses distributional correlation 
functions which have contact term ambiguities one should ask if the formula 
itself is free from such ambiguities. The contact term ambiguities arise from 
the choice of renormalization  scheme  and are generated by adding to the 
generating functional finite local counterterms of the form 
\begin{equation}\label{contact_redef}
 \ln Z[\lambda, g_{ij}] \mapsto \ln Z[\lambda, g_{ij}] + \int d^2x [ f(\lambda) \mu^{2}R_{2}(x) + 
\frac{1}{2}c_{ij}(\lambda) \partial_{\mu}\lambda \partial^{\mu}\lambda(x) + \dots] 
\end{equation}
where $f(\lambda)$ and $c_{ij}(\lambda)$ are arbitrary 
functions\footnote{We assume that these functions are at least once differentiable.} (scalar and tensor respectively) 
and the omitted terms contain higher order derivatives of the metric and sources. The redefinition 
(\ref{contact_redef}) shifts the terms in the renormalization operation ${\cal D}(x)$. 
The low order terms shift  as 
\begin{eqnarray}
C(x) &\mapsto & C(x) + \beta^{i}\partial_{i}f(x)\, , \\
W_{i}(x) &\mapsto & W_{i}(x) -\partial_{i}f(x)  - c_{ij}\beta^{j}(x) \, , \\
G_{ij}(x) & \mapsto & G_{ij}(x) - {\cal L}_{\beta}c_{ij}(x) \, . 
\end{eqnarray} 
with all shifts proportional to the identity operator.

The $c$-function and the metric tensors $g_{ij}$, $\Delta g_{ij}$ can  each be written in a form 
involving two point correlators at non-zero  separation only (see formulas (\ref{eq:corig}), (\ref{eq:gijorig}), 
(\ref{eq:gLij})). 
Thus these quantities are independent of the contact term ambiguities. The 1-form $w_{i}$ defined 
in (48) changes under (\ref{contact_redef}) as
\begin{equation}
w_{i} \mapsto w_{i} - \partial_{i}f 
\end{equation} 
and the antisymmetric form $b_{ij}$ thus does not change. Since the redefinition (\ref{contact_redef}) is 
the most general one\footnote{The higher order terms omitted in (\ref{contact_redef}) do not contribute to the change 
of $w_{i}$.} the two-form $b_{ij}$ is also independent of the contact term ambiguities.  


Another property that we would like to check is whether the quantities we defined depend on the 
scales $\mu$ and $\Lambda$ only via their ratio $\mu/\Lambda$. For the $c$-function (\ref{eq:corig}), the metric (\ref{eq:gijorig}) 
and the antisymmetric form (\ref{eq:bijorig})  this immediately
follows from the scaling properties (\ref{eq:muscale}). As for the metric correction $\Delta g_{ij}$ it may happen 
that the infrared regulated quantity $\Delta g_{ij}^{L}$ contains a logarithmic divergence $\sim \ln L$ whose 
subtraction requires introducing a new scale. If this happens the subtracted correction will not depend on 
$\mu$ and $\Lambda$ via the ratio $\mu/\Lambda$ only. The physical significance of this is unclear to us.


\subsection{The infrared condition: an example}\label{IRexample}

Here we discuss a simple example that demonstrates the necessity of
the infrared condition (\ref{eq:IRcond}) 
for a gradient formula to hold. Consider a free compact boson $X$
defined  on 
a two-dimensional curved surface with metric $g_{\mu\nu}$ by the
action functional 
\begin{equation}
S[R, g_{\mu\nu}]=\frac{1}{8\pi}\int\!\! d^{2}x\,
(\lambda\sqrt{g}g^{\mu\nu}\partial_{\mu}X\partial_{\nu}X 
+ QX\sqrt{g}R_{2}) 
\end{equation}
where $\lambda$ is the coupling constant corresponding to the radius
of compactification squared, 
$R_{2}$ is the curvature of $g_{\mu\nu}$, and $Q$ is a parameter. 
Promoting $\lambda$  to a local source $\lambda(x)$ we can define a
generating functional 
\begin{equation} \label{Zfunc}
\ln Z[\lambda(x), g_{\mu\nu}(x)]= \int\! [dX]\, e^{-S[ \lambda(x),
g_{\mu\nu}(x)]}\, .
\end{equation}
For the zero mode integral to be well defined we assume that the
theory is defined only on a surface with 
the topology of a plane so that 
\begin{equation}
\int\!\!d^2 x\,  \sqrt{g}R_{2} = 0 \, . 
\end{equation}
and the zero mode integral in (\ref{Zfunc}) only yields an overall numerical factor.
Note that $Q$  cannot be considered as  a coupling constant as it
does not stand at a local operator. The functional integral is
Gaussian so the anomaly 
can be readily computed (e.g. using the heat kernel method) with the
result 
\begin{equation}
D(x)=\Theta(x) = \frac{1}{2}C(\lambda)\sqrt{g}R_{2}(x) +
J_{\lambda}^{\mu}(x)\partial_{\mu}\lambda +
\frac{1}{2}g_{\lambda\lambda}\partial_{\mu}\lambda\partial^{\mu}\lambda
+ \partial_{\mu}(w_{\lambda}\partial^{\mu}\lambda)
\end{equation}
where 
\begin{eqnarray}\label{Dcoefs}
&& C(\lambda) = \frac{1}{12\pi} + \frac{Q^2}{4\pi \lambda} \\
&& J_{\lambda}^{\mu}(x)= -\frac{Q}{4\pi \lambda}\partial^{\mu} X(x)\\
&& g_{\lambda\lambda} =  \frac{1}{64\pi \lambda^2} 
\end{eqnarray}
The value of $w_{\lambda}$ is essentially scheme dependent. It can be
shifted by adding to $S$ a local 
counterterm $\int\! d^2 x \, f(\lambda(x))R_{2}(x)$ dependent on an
arbitrary function $f(\lambda)$. 
In the context of nonlinear sigma models 
such term can be fixed by target space diffeomorphism invariance. For
the model at hand this gives 
$w_{\lambda}=  (8\pi\lambda)^{-1}  $.    

We see from (\ref{Dcoefs}) that while the theory has a vanishing
beta function, its $c$-function: 
$c=12{\pi}C(\lambda)$ has a nontrivial 
derivative with respect to  the modulus $\lambda$. 
We can further observe that it is the broken global conformal
symmetry that is 
responsible for the breakdown of gradient property. The stress-energy
tensor on a flat surface is 
\begin{equation}
T_{\mu\nu}= \frac{\lambda}{4\pi}(:\partial_{\mu} X\partial_{\nu}X:
-\frac{\delta_{\mu\nu}}{2}
:\partial_{\gamma}X\partial^{\gamma}X:) 
 + \frac{Q}{4\pi}(\delta_{\mu\nu}\partial_{\lambda}\partial^{\lambda}
- \partial_{\mu}\partial_{\nu})X 
\end{equation} 
It has exactly the same form as the background charge model
\cite{Dotsenko}  with imaginary background charge. Note  that 
in our theory there is no background charge. Moreover since our
theory is defined on a topological plane the field $X$ can be taken 
to be compact with an arbitrary radius.  The
correlation function 
\begin{equation}
\langle T(z) J_{\lambda, z}(0) \rangle = -\frac{Q^{2}}{4\pi
\lambda^2}\frac{1}{z^{3}}
\end{equation}
means that special conformal transformations are broken by the
boundary condition at infinity,
\footnote{ The charge $\oint dz\, z^{2} T(z)$ does not vanish at
infinity.}.

Another way to see the necessity to have a theory defined on a sphere
of large radius is in 
the context of nonlinear sigma model. There it is essential for the
gradient formula to hold 
(at least in the leading order in the $\alpha'$ expansion) that the
zero mode measure includes the dilaton contribution corresponding to
spherical topology \cite{Osborn}. 

\subsection{Bare gradient formula} \label{sect:baregrad}
Here we will show how the Wess-Zumino consistency condition for the local renormalization 
operation can be used to derive a different gradient formula. The main quantities in the new 
gradient formula are constructed using the anomalous contact terms present in $\cal D$ 
rather than correlation functions 
at finite separation. For this reason we call it a bare gradient formula. 
As a consequence of that the terms in that formula are defined modulo contact term
 ambiguities discussed in section \ref{contterms}. The new formula also suffers from potential infrared divergences 
 in the metric. 
 In this section however for the sake of brevity we will not introduce an explicit infrared cutoff and our 
 manipulations with integrals will be  formal. It is straightforward however to introduce such a cutoff 
 with the main result correct up to some error terms vanishing when the cutoff is removed.  

Using (\ref{eq:3comms}) the  Wess-Zumino consistency condition 
\begin{equation} \label{eq:WZgen}
[(D(x_2)- {\cal D}(x_2) ), (D(x_1) - {\cal D}(x_1)) ] \, \rangle \rangle = 0 
\end{equation}
 can be rewritten as\footnote{Note that this form of the Wess-Zumino condition is linear in 
 ${\cal D}$. This leads to  essential simplifications in computations and also ensures that terms 
 with tensorial sources in ${\cal D}$ do not contribute to the final gradient formula.}   
\begin{eqnarray}
&& \Bigl[ [\Theta(x_2), {\cal D}(x_1)] - [\Theta(x_1), {\cal D}(x_2)] - [\beta(x_2), {\cal D}(x_1)] 
\nonumber \\
&& + \beta(x_1)D(x_2) - {\cal D}(x_2) \Theta(x_1) + {\cal D}(x_1)D(x_2) \Bigr]\, \rangle \rangle =0 \, . 
\end{eqnarray}
Applying to the above equation $\langle \langle \phi_{i}(y)$ on the left and integrating over $x_{1}$ we obtain 
\begin{eqnarray}\label{eq:WZnext}
&& \langle \langle \phi_{i}(y)[ {\cal D}\Theta(x_2) - \beta^{j}{\cal D}\phi_{j}]\rangle\rangle 
+ \langle\langle \phi_{i}(y) \int\!\! d^{2} x_1\, \beta(x_1) D(x_2)\rangle \rangle  
+ \langle \langle {\cal D} \phi_{i}(y) D(x_2) \rangle \rangle \nonumber \\
&& -\mu \frac{\partial}{\partial \mu} ( \langle D(x_2) \phi_{i}(y)\rangle_{c} 
- \delta^{2}(y-x_2)\partial_{i}\beta^{j}\langle \phi_{j}\rangle) = 0
\end{eqnarray}
where we used the identities 
\begin{equation}
\int\!\! dx_1 [\Theta(x_1), {\cal D}(x_2)] = 0 \, , \quad 
\langle \langle \phi_i(y) \int\!\! dx_1 [\beta^{j}(x_2),  {\cal D}(x_1) ] \phi_{j}(x_2)\rangle\rangle = 0\, .
\end{equation}

As we know from section \ref{sect:DTheta}
$
{\cal D}\Theta - \beta^{j}{\cal D}\phi_{j}$ is a pure-contact operation. Its field part
 $ \beta^{j}\partial_{\mu}J^{\mu}_{i} - \partial_{\mu}J^{\mu}$ vanishes (is pure contact). 
 Equation (\ref{eq:WZnext}) expresses the contact terms with $\phi_{i}(y)$ via the operation $\cal D$.  
Integrating the above formula over $x_2$ with the weight $(x_2-y)^{2}$ and using 
\begin{equation}
\mu \frac{\partial}{\partial \mu} \int\!\! d^{2}x_2\, \langle D(x_{2})\phi_{i}(y) \rangle_{c} (x_2-y)^{2} = 0
\end{equation}
we obtain\footnote{Recall that the currents $J_{i}^{\mu}$ and the metric $G_{ij}$ in 
(\ref{eq:Doploose}) are ordinary operations so that $\langle \partial_{i}J_{j}^{\mu}\rangle = 0$.} 
\begin{equation}
\partial_{i}\langle C_{2} \rangle = - H_{ij}\beta^{j} + {\cal L}_{\beta}W_{i} + Q_{i}
\end{equation}
where 
\begin{eqnarray}
H_{ij} &=& -G_{ij} 
- \frac{1}{4} \int\!\! d^{2} y \, y^{2} [ \langle  \partial_{\mu}J^{\mu}_{j}(0) \phi_{i}(y)  \rangle 
+ \langle \partial_{\mu}J^{\mu}_{i}(0)  \phi_{j}(y) \rangle  \,] \,  , \nonumber \\
G_{ij}&=& - \frac{1}{4} \int\!\! d^{2} y \, y^{2} \langle \langle [\phi_{i}(0), {\cal D}\phi_{j}(y)]\, 
\rangle \rangle \, , \nonumber \\
W_i &=&  \frac{1}{4} \int\!\! d^{2} y \, y^{2} \langle D(y) \phi_{i}(0)\rangle_{c} \, , \nonumber \\
Q_{i} &=& \frac{1}{4}\int\!\! d^2 y\,  y^{2} \langle \partial_{\mu}J^{\mu}_{i}(y) \Theta(0)\rangle_{c} \, . 
\end{eqnarray}
Note that the tensor $G_{ij}$ is symmetric. This follows from the fact that operations $\phi_{i}(y)$, 
$\phi_{j}(x)$ commute.   The metric tensor $H_{ij}$ can be also written in terms of integrated 
correlation functions 
\eq
H_{ij} = \frac{1}{4}\int\!\! d^{2} y \, y^{2} \Bigl[ \int\!\! d^{2}x\, 
\langle D(x)\phi_{i}(y)\phi_{j}(0)\rangle_{c} - \partial_{i}\beta^{k}\langle \phi_{k}(y)\phi_{j}(0)\rangle_c 
- \partial_{j}\beta^{k}\langle \phi_{i}(y)\phi_{k}(0)\rangle_{c} \Bigr] \, .  
\en

 According to our main infrared assumption (\ref{eq:IRcond}) $Q_{i}$ vanishes and we have a gradient formula
 \begin{equation} \label{eq:baregrad}
 \partial_{i} c^{(0)} + g_{ij}^{(0)}\beta^{j} + b_{ij}^{(0)}\beta^{j} = 0 
 \end{equation} 
where 
\begin{equation}
c^{(0)}= \langle C_{2}\rangle - W_{i}\beta^{i}\, , \quad g_{ij}^{(0)}= H_{ij}\, , \quad 
b_{ij}^{(0)}= \partial_{i}W_{j} - \partial_{j}W_{i} \,. 
\end{equation}
The metric $H_{ij}$ potentially suffers from the  same infrared divergences as the correction to Zamolodchikov's 
metric defined in (\ref{eq:deltagij}). We define the finite quantity entering (\ref{eq:baregrad}) 
by subtracting these divergences.

When loose power counting applies the above quantities can be computed more explicitly using 
(\ref{eq:Doploose}). In this case we have 
\begin{equation}
H_{ij} =  G_{ij} - \frac{1}{4}\int\!\! d^{2}y \, y^{2} [\langle \partial_{\mu}J^{\mu}_{i}(y) \phi_{j}(0) 
\rangle + \langle \partial_{\mu}J^{\mu}_{j}(y) \phi_{i}(0) 
\rangle] \, , 
\end{equation}
 $\langle C_{2}\rangle = \langle C \rangle$ and $G_{ij}$, $W_{i}$ coincide with the respective 
 quantities  defined in formula 
(\ref{eq:Doploose}). 
In the case when the currents $J_{i}^{\mu}$  are absent formula 
(\ref{eq:baregrad}) matches with the one obtained by Osborn \cite{Osborn}.

 \subsection{Dressing transformations} 

For any gradient formula 
\begin{equation}  
 \partial_{i}c +  g_{ij}\beta^{j} + b_{ij} \beta^{j}=0
\end{equation} 
with a symmetric tensor $g_{ij}$ and an antisymmetric tensor  $b_{ij}$ 
 one can redefine $c$, $b_{ij}$ and $g_{ij}$ as 
 \begin{eqnarray}\label{eq:dressing}
 \tilde c &=& c + \beta^{i}c_{ij}\beta^{j} \, , \nonumber \\
 \tilde g_{ij} &= &g_{ij} - {\cal L}_{\beta} c_{ij} \, , \nonumber \\
 \tilde b_{ij} &=& b_{ij} - (d i_{\beta} c)_{ij}
 \end{eqnarray}
 so that a gradient formula $\partial_{i}\tilde c =   \tilde g_{ij}\beta^{j} +\tilde b_{ij} \beta^{j}$ holds. 
 The tensor $c_{ij}$ above is any tensor on the space of couplings that may depend on the couplings and the 
 renormalization scale $\mu$. We will refer to redefinitions (\ref{eq:dressing}) as dressing transformations.
 One can show that formula (\ref{eq:finalgradf}) is related to formula (\ref{eq:baregrad}) by means of a dressing 
 transformation specified by 
 \begin{equation}
 c_{ij}^{\Lambda} = \int\!\! d^{2}x\, G_{\Lambda}(x) \langle \phi_{i}(x) \phi_{j}(0)\rangle_{c}
 \end{equation}
 so that 
 \begin{equation}
 c= c^{(0)} - \beta^{i}c_{ij}^{\Lambda}\beta^{j} \, . 
 \end{equation}
 
 It is not hard to construct using dressing transformations a class of c-functions that monotonically 
 decrease under the RG flow. Such functions $c^{f}$ can be defined as 
 \begin{equation}
 c^{f} = -3\pi \int\!\! d^{2}x\, x^{2}f(x^{2})\langle \Theta(x) \Theta(0)\rangle_{c} 
 \end{equation}
 where $f(x^2)$ is a function such that $f(0)=1$, $f(x^{2})$ decreases fast at infinity\footnote{An exponential decrease 
 would suffice for all purposes.} and 
 \begin{equation}
 x^{\mu}\partial_{\mu} f(x^{2}) < 0 \, . 
 \end{equation}  
 These potential functions satisfy a gradient formula 
 \begin{equation}
 \partial_{i}c^{f} = -(g_{ij}^{f} + \Delta g_{ij}^{f} + b_{ij}^{f} )\beta^{j} 
 \end{equation}
 where
 \begin{equation}
 g_{ij}^{f}= -3\pi \int\!\! d^{2}x\, x^{2} [x^{\mu}\partial_{\mu} f(x^{2})] \langle \phi_{i}(x)\phi_{j}(0)\rangle_{c}
 \end{equation}
 \begin{equation}
 \Delta g_{ij}^{f} = 3\pi \int\!\! d^{2}x\, x^{2}[f(x^{2})-1]
 (\langle \partial_{\mu}J^{\mu}_{i}(x)\phi_{j}(0)\rangle + \langle \partial_{\mu}J^{\mu}_{j}(x)\phi_{i}(0)\rangle)
 \end{equation}
 \begin{equation}
b_{ij}^{f} = \partial_{i}w_{j}^{f} -  \partial_{j}w_{i}^{f}
 \end{equation}
 \begin{equation}
 w_{i}^{f}= 3\pi \int\!\! d^{2}x\, x^{2} f(x^{2}) \langle \phi_{i}(x)\Theta(0)\rangle_{c}  
 \end{equation}
 
 Such smeared $c$-functions were first considered in \cite{cspec}.
 

\subsection{Renormalization group transformation as a flow of couplings} 
As one can observe from the form of Callan-Symanzik equations (\ref{eq:CS3}) 
the scale transformation of correlation functions 
$$
\langle \phi_{i_1}(x_1)\phi_{i_2}(x_2) \dots \Theta(y_1)\Theta(y_2) \dots \rangle_{c} 
$$ even at finite separation is not fully 
compensated by the change of couplings $\lambda^i$. In addition to changing the couplings according 
to their beta functions and rotating the fields $\phi_{i}$ by the anomalous 
dimension matrices $\partial_{i}\beta^{j}$ the operators $\phi_{i}(x)$ and $\Theta(y)$ each shift 
by an additional total derivative: $\partial_{\mu}J^{\mu}_{i}(x)$ and  $\partial_{\mu}J^{\mu}(y)$ 
respectively. If the currents $J^{\mu}_{i}$, $J^{\mu}$ are not conserved these shifts  affect 
the scale transformation of the correlation functions taken at finite separation. This signals 
that more couplings need to be introduced to parameterize such additional terms in the Callan-Symanzik 
equations. Thus to account for the current $J^{\mu}(y)$ it is customary  to introduce 
dilaton couplings $\lambda^{i}_{D}$ that couple to $\phi_{i}(x)\mu^{2}R_{2}(x)$ terms in the Lagrangian
\footnote{A completeness of the set $\phi_{i}$ is assumed here as discussed in section 5.}. 
The generating functional $Z$ depends on these couplings according to the functional differential equation 
\begin{equation}\label{eq:dil}
\frac{\delta \ln Z}{\delta \lambda^{i}_{D}(x)}  = 
\frac{1}{2}\mu^{2}R_{2}(x)\frac{\delta \ln Z}{\delta \lambda^{i}(x)} \, . 
\end{equation} 

The introduction of this new set of couplings is natural if one bears in mind 
that coupling constant redefinitions are responsible for having different RG schemes. 
To renormalize a theory on a curved space one needs counterterms of the form $\phi_{i}(x)\mu^{2}R_{2}(x)$. 
As usual such counterterms are defined up to arbitrary finite parts. Changing the dilaton 
couplings $\lambda^{i}_{D}$ accounts for changing the finite parts in such counterterms. 
(Previously we assumed that  such counterterms are fixed somehow which amounts to partially fixing 
the RG scheme. This resulted in the extra terms in the Callan-Symanzik equations.) 
Expanding the operator $C(x)$ in (\ref{eq:Doploose}) as $C(x) = \beta^{i}_{D}\phi_{i}(x)$ we 
see that the coefficients $\beta^{i}_{D}$ can now be naturally interpreted as the beta functions 
for the dilaton couplings. 
For the loose power counting case the Callan-Symanzik equation for correlators of stress-energy tensor 
takes the form 
\begin{eqnarray}\label{eq:CSwithdil}
&&\mu\frac{\partial}{\partial \mu} \langle T_{\mu \nu}(y_1) T_{\alpha \beta}(y_2) \dots \rangle_{c} 
= \beta^{i}\frac{\partial}{\partial \lambda^{i}}\langle T_{\mu \nu}(y_1) T_{\alpha \beta}(y_2) \dots \rangle_{c} 
+ \langle  \Gamma_{\mu\nu}^{C}(y_1)T_{\alpha \beta}(y_2) \dots \rangle_{c} \nonumber \\ 
&& + \langle  T_{\mu \nu}(y_1) \Gamma_{\alpha \beta}^{C}(y_2) \dots \rangle_{c} + \dots 
=(\beta^{i}\frac{\partial}{\partial \lambda^{i}} + \beta^{i}_{D}\frac{\partial}{\partial \lambda^{i}_{D}})\langle T_{\mu \nu}(y_1) T_{\alpha \beta}(y_2) \dots \rangle_{c} 
\end{eqnarray}
where 
\begin{equation}
\Gamma_{\mu \nu}^{C}(x) = (\partial_{\mu}\partial_{\nu} - g_{\mu \nu}\partial_{\alpha}\partial^{\alpha})C(x) \, . 
\end{equation}
We used (\ref{eq:dil}) and (\ref{eq:Doploose}) to obtain the last equality in (\ref{eq:CSwithdil}).
We see that the dilaton couplings account for mixings of the stress energy tensor with trivially conserved 
currents $\Gamma_{\mu \nu}^{C}(x)$. With the enlarged set of couplings $(\lambda, \lambda_{D})$ 
the change in scale for correlators of stress-energy tensor components (at finite separation) 
is exactly compensated by the change in coupling constants. In particular for the $c$-function 
(\ref{eq:corig}) we have 
\begin{equation}\label{eq:ccov}
\mu \frac{\partial c}{\partial \mu} = (\beta^{i}\frac{\partial}{\partial \lambda^{i}} + \beta^{i}_{D}\frac{\partial}{\partial \lambda^{i}_{D}}) c\, . 
\end{equation}
We can also compute the derivatives of the 
$c$-function (\ref{eq:corig}) with respect to the dilaton couplings. Using (\ref{eq:c}), 
(\ref{eq:dil}) and the identity 
\begin{equation}
\frac{\partial}{\partial \lambda^{i}_{D}} = \int\!\! d^{2}x \, \frac{\delta}{\delta \lambda^{i}_{D}(x)}
\end{equation} 
we obtain 
\begin{equation}\label{eq:partialc2} 
\frac{\partial c}{\partial \lambda^{i}_{D}} = -\frac{\partial}{\partial \lambda^{i}_{D}}
\int\!\! d^2 x\, G_{\Lambda}(x) \langle \Theta(x) \Theta(0)\rangle_{c}  = 
2 \int\!\! d^2 x\, G_{\Lambda}(x) \langle \Theta(x) \partial_{\mu}\partial^{\mu}\phi_{i}(0)\rangle_{c} \, . 
\end{equation}
Integrating by parts in (\ref{eq:partialc2}),  using 
\begin{equation}
\mu \frac{\partial}{\partial  \mu} \langle \phi_i\rangle = \beta^{j} \partial_{j} \langle \phi_i\rangle 
\end{equation}
and the assumption that 
\begin{equation}
\int\!\! d^{2} x \, \langle \phi_{j}(x) \phi_{i}(0)\rangle = \partial_{j}\langle \phi_{i}\rangle 
= \partial_{i} \langle \phi_j \rangle < \infty \, , 
\end{equation}
we obtain 
 \begin{equation} \label{eq:dilgrad}
  \frac{\partial c}{\partial \lambda^{i}_{D}} = -g_{ij}^{D}\beta^{j}
\end{equation}
where 
\begin{equation} \label{eq:dilmetric}
g_{ij}^{D} = 2\int\!\!d^{2}x\, [G_{0}(x) - G_{\Lambda}(x)]\langle \phi_{j}(x) \partial_{\mu}\partial^{\mu}\phi_{i}(0)\rangle_{c} 
\end{equation}
is a  symmetric tensor.

We can further show that the contraction of gradient formula (\ref{eq:finalgradf}) with the beta functions $\beta^{i}$ gives Zamolodchikov's 
formula (\ref{eq:zam}).  This boils down to the identity 
\begin{equation} \label{eq:beta_contr}
\beta^{i}_{D}\frac{\partial c}{\partial \lambda_{D}^{i}} = \beta^{i}\Delta g_{ij}\beta^{j}\, . 
\end{equation} 
Using equations (\ref{eq:dilgrad}), (\ref{eq:dilmetric}) the left hand side of 
equation  (\ref{eq:beta_contr}) can be written as 
\begin{equation} \label{eq:beta_contr2}
\beta^{i}_{D}\frac{\partial c}{\partial \lambda_{D}^{i}} = 2\int\!\!d^{2}x\, [G_{\Lambda}(x) - G_{0}(x)]
\langle \Theta(x) \partial_{\mu}\partial^{\mu}C(0)\rangle_{c} 
\end{equation}
while for the right hand side we have 
\begin{equation} \label{eq:beta_contr3}
\beta^{i}\Delta g_{ij}\beta^{j}=2\int\!\!d^{2}x\, [G_{\Lambda}(x) - G_{0}(x)]
\langle \Theta(x)\beta^{i} \partial_{\mu} J^{\mu}_{i}(0)\rangle_{c} \, .
\end{equation}
The last expression coincides with (\ref{eq:beta_contr2}) by virtue of the identity $\partial_{\mu}\partial^{\mu}C(x) =\partial_{\mu} J^{\mu}_{i}(x)$
proven in section \ref{sect:DTheta}. This identity can be used because the two point function in (\ref{eq:beta_contr3}) is taken at finite separation. 
It is not hard to extend the proof of identity (\ref{eq:beta_contr}) to a more general case not assuming the loose power counting. Formula 
(\ref{eq:beta_contr}) shows in particular that the metric correction $\Delta g_{ij}$ is necessary to account for the flow of dilaton coupling constants 
when the last ones are present.

The additional gradient formula (\ref{eq:dilgrad}) together with the main formula (\ref{eq:finalgradf}) imply that 
the $c$-function is stationary with respect to the couplings $(\lambda, \lambda_{D})$ at fixed points 
$\beta^{i}=0$. The inverse follows from the Zamolodchikov's formula (\ref{eq:zam}) combined with 
formula (\ref{eq:ccov}). Thus under our main set of assumptions and 
 with loose power counting the stationary points of the $c$-function are in a one to one 
correspondence with the fixed points.

\section{Final comments}\label{sec:final} 
As we said in the introduction one of the motivations to obtain  a general gradient formula came from string theory. 
In regard with potential applications of  our result to the problem of constructing string effective 
actions  it should be stressed  that we worked throughout with normalized connected correlation functions 
while it is the unnormalized and disconnected ones which are relevant to string theory. This fact also 
explains why our results {\it seem} to be at odds with the conclusion of \cite{Tseytlin_c} that the 
Zamolodchikov $c$-function  does not give a suitable string effective action. In the unnormalized correlators 
the dilaton zero mode $\phi_{0}$ contributes an overall factor $e^{-2\phi_{0}}$ which results in having the same 
factor in $c$. Thus stationarity of $c$ with respect to $\phi_{0}$ implies that $c$ has to vanish at stationary 
points. This factor and the related problem disappear when one builds $c$ out of normalized correlators 
as we do in this paper. 

The aforementioned problem with $c$ prompted various authors to switch to using what we call the bare  
gradient formula which was discussed in section 9.3. The negative side of this is that the  metric that appear 
in that formula, being built from contact terms, does not have any positivity properties.  

In the present paper we focused on a formal derivation of the new gradient formula and discussing its general properties. 
It would be instructive to illustrate how it works on concrete examples in conformal perturbation theory and nonlinear 
sigma models. We are planning to do this in a separate publication \cite{FKinprep}. It is also interesting to understand 
better the implications of the new formula for string theory. We leave this question to future studies.

\begin{center}
{\bf \large Acknowledgments}
\end{center}
The work of D.F. was supported by the Rutgers New High Energy Theory Center. Both authors  acknowledge the support 
of   Edinburgh Mathematical Society.

\bibliographystyle{plain}
 
\end{document}